\documentclass[amsmath,aps,pra,reprint,notitlepage,twocolumn,10pt,superscriptaddress,floatfix,nofootinbib]{revtex4-1}

\usepackage{graphicx}
\usepackage{bm}
\usepackage{hyperref}
\usepackage{bbm}
\usepackage{color}
\usepackage{multirow}

\begin{document}
\title{Screening and enhancement of oscillating electric field in molecules}
\author{H. B. Tran Tan}
\affiliation{School of Physics, University of New South Wales, Sydney, New South Wales 2052, Australia}
\author{V. V. Flambaum}
\affiliation{School of Physics, University of New South Wales, Sydney, New South Wales 2052, Australia}
\affiliation{Helmholtz Institute Mainz, Johannes Gutenberg University, 55099 Mainz, Germany}
\author{I. B. Samsonov}
\affiliation{School of Physics, University of New South Wales, Sydney, New South Wales 2052, Australia}

\date{\today}
\begin{abstract}
\begin{center}
\textbf{Abstract}\\
\end{center}
According to the Schiff theorem, the atomic electrons completely screen the atomic nucleus from an external static electric field. However, this is not the case if the field is time-dependent. 
Electronic orbitals in atoms either shield the nucleus from an oscillating electric field when the frequency of the field is off the atomic resonances or enhance this field when its frequency approaches an atomic transition energy. In molecules, not only electronic, but also rotational and vibrational states are responsible for the screening of oscillating electric fields. As will be shown in this paper, the screening of a low-frequency field inside molecules is much weaker than it appears in atoms owing to the molecular ro-vibrational states. We systematically study the screening of oscillating electric fields inside diatomic molecules in different frequency regimes,i.e., when the field's frequency is either of order of ro-vibrational or electronic transition frequencies. In the resonance case, we demonstrate that the microwave-frequency electric field may be enhanced up to six orders in magnitude due to ro-vibrational states. We also derive the general formulae for the screening and resonance enhancement of oscillating electric field in polyatomic molecules. Possible applications of these results include nuclear electric dipole moment measurements and stimulation of nuclear reactions by laser light.

\end{abstract}

\maketitle

\section{Introduction}

It is well-known that the Standard Model of elementary particles
predicts small electric dipole moments (EDMs) for the electron and nucleon, see, e.g., Refs.\ \cite{Chupp,Yamanaka,Safronova} for recent reviews. Different extensions
of SM such as axion or supersymmetry predict, however, different
values for the EDMs of elementary particles. Therefore, it is a
challenge for modern experimental physics to measure the EDMs of
the electron and neutron in order to verify (or falsify) these models. Presently, there are several groups pursuing this goal, although the sensitivity of current experiments does not allow one to make firm conclusions.

One of the difficulties encountered in measuring the EDMs of nuclei in atoms and molecules is the screening of the external electric field by the electron shells in
these systems. Indeed, according to a theorem by Schiff\ \cite{Schiff}, the atomic
nucleus of a neutral atom is completely screened from any \textit{static}
external electric field. EDMs of diamagnetic atoms are produced by the interaction of electrons with the nuclear Schiff moment \cite{Sandars1,Sandars2,SFK1984,FLAMBAUM1985,FLAMBAUM1986}. For light atoms, atomic EDMs produced by the Schiff moment are very small, while they appear significant for heavy
atomic species ($\sim 10^{-3}d_n$; $d_n$ is the EDM of neutron). Thus, the electron screening
makes the detection of the nuclear EDMs difficult.

A possible way to overcome this difficulty is to use ions
instead of neutral atoms, where the screening of external fields
is incomplete\ \cite{FlambaumConstant}. However, since a charged
particle is not stationary in an electric field, it is problematic
to preform precise measurements on such particle.

The behavior of atoms in an {\it oscillating} electric field is
drastically different from that in the static case.
It is natural to expect that the screening of alternating electric
field in atoms and molecules is incomplete since the particles constituting an atom or a molecule respond to the changes in the field with some delay. Recently, it has
been shown that when the frequency of the external electric field
is far from atomic resonances, the resulting electric field at the
center of an atom is proportional to the dynamical atomic
polarizability \cite{FlambaumOscillating}. Numerical tests of these results were performed in Ref.\ \cite{DBGF}. However, when the
frequency of the external field approaches the energy of atomic
transition, there may be a significant enhancement (up to $10^5$)
of the field\ \cite{SamsonovResonant}. 

%

The extension of the static Schiff theorem to molecules was
considered in Ref.\ \cite{FlambaumMolecule}. The derivation therein
used the Ehrenfest theorem and resorted to classical mechanics to
relate the acceleration of each nucleus to that of the whole
molecule. Although this approach proved fruitful in the static
case, it becomes inefficient in the dynamic case since the
classical motion of a molecule in an oscillating field is itself
difficult to describe. This paper aims at developing a general and
fully quantum mechanical method for computing the electric field inside atoms and molecules which is
applicable both for static and oscillating electric
fields.

Note that the quantum mechanical description of molecules has some
important features as compared with the atomic case. In molecules,
it is necessary to separate the relative motion of the
constituent nuclei from the motion of the common center of mass while
in atoms this separation is not essential, i.e., the atomic nucleus may be considered as fixed in space.
This relative
motion of the nuclei in a molecule is described by
the rotational and vibrational modes. As a result, the molecular
spectra appear to have a very rich structure with rotational,
vibrational and electronic states. As will be shown in this paper,
these states give rise to new terms in the formula for the
resulting electric field in the molecule as compared with the
atomic case studied in \cite{FlambaumOscillating}. All these new terms
play an important role in the screening of the external
electric field in molecules.

The rest of this paper is organized as follows. In Sect.\
\ref{Diatomic_Theoretical}, the problem of calculating the field
at the nuclei in a diatomic molecule in an external electric field
is discussed. The cases of a static field, of an oscillating field
off resonance with the molecular transitions and of an oscillating field
on resonance with a molecular transition will be examined. In
Sect.\ \ref{Diatomic_Numerical}, we will consider examples of some diatomic molecules and present the estimates of screening and resonance enhancement of electric field at nuclei.
Sect.\ \ref{General Consideration} is
devoted to the generalization of the diatomic-molecule case to
systems of arbitrary number of nuclei. Sect.\ \ref{Conclusion}
contains a summary of the results and discussion of their potential application. 


Throughout this paper, we employ the atomic units in which $\hbar =
e = m_e = 4\pi\epsilon_0= 1$. This makes the intermediate
formulae more compact. The final results, for
convenience, will be presented in arbitrary units with all
fundamental constants given explicitly.

\section{Screening of electric field in diatomic molecules}\label{Diatomic_Theoretical}

In this section, we derive the general formulae for the screening of both static and alternating external electric fields inside diatomic molecules. To make our presentation self-consistent, we begin with a review of the molecular Hamiltonian in the center-of-mass coordinates frame.



\subsection{The diatomic molecule Hamiltonian in the center-of-mass frame}\label{SecIIA}

Consider a diatomic molecule with $L$ electrons in an external electric
field $\mathbf{E}_{\rm ext}$. Let $M_I$ and $Z_I$ be the masses
and charges of the nuclei, respectively (the subscript $I=1,2$
labels the nuclei). The position and momentum operators of the
nuclei in the laboratory frame will be denoted by $\mathbf{R}_I$ and $\mathbf{P}_I$, respectively.
The electron positions and momenta operators will be denoted by $\mathbf{r}_i$ and $\mathbf{p}_i$ (the subscript $ i=1,\ldots,L$ labels the electrons).

The Hamiltonian of the diatomic molecule in the laboratory
frame has the standard form
\begin{subequations}\label{Diatomic_Hamiltonian}
\begin{eqnarray}
  H_{\rm mol}&=&K+V_{0}+V \,,\\
  K&=&\frac{\mathbf{P}_1^2}{2M_1}+\frac{\mathbf{P}_2^2}{2M_2}
   +\sum\limits_{i=1}^{L}{\frac{\mathbf{p}_i^2}{2}}\,,\\
  V_0&=&{\frac{{{Z}_{1}}{{Z}_{2}}}{{{R}_{12}}}} -\sum\limits_{i=1}^{L}
   \left(\frac{Z_1}{R_{1i}}+\frac{Z_2}{R_{2i}}\right)
   +\sum_{i< j}^{L}\frac1{r_{ij}}\,,\label{1c}\\
  V&=&-{{\mathbf{E}}_{\rm ext}}\cdot \left(Z_1\mathbf{R}_1+Z_2\mathbf{R}_2
   -\sum\limits_{i=1}^{L}\mathbf{r}_{i} \right) \,,
\end{eqnarray}
\end{subequations}
where $R_{Ii}=\left| \mathbf{R}_I-\mathbf{r}_i \right|$,
$r_{ij}=\left| \mathbf{r}_i-\mathbf{r}_{j} \right|$ and
$R_{IJ}=\left| \mathbf{R}_{I}-\mathbf{R}_J \right|$.
Recall that we are using
the atomic unit system in which $e=m_e=1$. Here, for simplicity, we consider the non-relativistic Hamiltonian for spinless particles.

It is convenient to define the total nuclear mass
$M_N=M_1+M_2$, the total nuclear charge $Z_N=Z_1+Z_2$,
the total molecular mass $M_T=M_N+L$ and the total
molecular charge $Z_T=Z_N-L$ ($Z_T=0$ for a neutral
molecule).

To separate the molecule's center-of-mass motion from the relative
dynamics of the electrons and nuclei, we perform the change of
variables $({\bf R}_I,{\bf r}_i)\to ({\bf S}_T,{\bf S},{\bf
s}_i)$
\begin{equation}\label{Diatomic_Coordinates_Transformation}
\begin{aligned}
  {{\mathbf{S}}_{T}}&=\frac{1}{{{M}_{T}}}\left(M_1\mathbf{R}_1+M_2\mathbf{R}_2
   + \sum\limits_{i=1}^{L}{{{\mathbf{r}}_{i}}} \right)\,,\\
   \mathbf{S}&=\mathbf{R}_1-\mathbf{R}_2\,,\\
   {{\mathbf{s}}_{i}}&={{\mathbf{r}}_{i}}-{\frac{M_1{{\mathbf{R}}_{1}}+M_2{{\mathbf{R}}_{2}}}{{{M}_{N}}}}\,.
\end{aligned}
\end{equation}
where ${\bf S}_T$ is the position operator of the molecular center of mass, $\bf S$ defines the molecular axis and ${\bf s}_i$ are the positions of the electrons
with respect to the nuclear center of mass.


The conjugated momenta of the coordinates $({\bf S}_T,{\bf S},{\bf
s}_i)$ will be denoted by $({\bf Q}_T,{\bf Q},{\bf q}_i)$.
These momenta are related to the original momenta $({\bf P}_I,{\bf
p}_i)$ via
\begin{equation}\label{Diatomic_Momenta_Transformation}
\begin{aligned}
  {{\mathbf{P}}_{I}}&=-\frac{{{M}_{I}}}{{{M}_{N}}}\sum\limits_{i=1}^{L}{{{\mathbf{q}}_{i}}}
  -(-1)^I\mathbf{Q}+\frac{{{M}_{I}}}{{{M}_{T}}}{{\mathbf{Q}}_{T}} \,,\\
  {{\mathbf{p}}_{i}}&={{\mathbf{q}}_{i}}+\frac{1}{{{M}_{T}}}{{\mathbf{Q}}_{T}}
  \,.
\end{aligned}
\end{equation}

The change of variables\ \eqref{Diatomic_Coordinates_Transformation} and\ \eqref{Diatomic_Momenta_Transformation} in
the Hamiltonian\ \eqref{Diatomic_Hamiltonian} allows us to isolate the
dynamics of the center-of-mass from the
relative motion, 
\begin{equation}
\label{4}
H_{\rm mol}= H_T + H_{\rm rel}\,,
\end{equation}
where
\begin{equation}\label{CM_Hamiltonian}
H_T =\frac{\mathbf{Q}_{T}^{2}}{2{{M}_{T}}}-{{Z}_{T}}{{\mathbf{E}}_{\rm ext}}\cdot {{\mathbf{S}}_{T}}
\end{equation}
is the center-of-mass Hamiltonian and $H_{\rm rel}$ is the
Hamiltonian describing the relative dynamics
\begin{subequations}
\label{Diatomic_Transformed_Hamiltonian}
\begin{eqnarray}
H_{\rm rel}&=&H_0+V_{\rm rel} \,,\\
 H_0&=&\sum\limits_{i=1}^{L}{\frac{\mathbf{q}_{i}^{2}}{2{{\mu }_{e}}}}
 +\sum\limits_{i< j}^{L}{ \frac{{{\mathbf{q}}_{i}}{{\mathbf{q}}_{j}}}{{{M}_{N}}}}+\frac{\mathbf{Q}^2}{2\mu_N} + V_0 \,,\label{6b}\\
 V_{\rm rel}&=&V+Z_T\mathbf{E}_{\rm ext}\cdot\mathbf{S}_T\equiv-\mathbf{d}\cdot\mathbf{E}_{\rm ext}\,.
 \label{6c}
\end{eqnarray}
\end{subequations}

Here
\begin{equation}
\label{reduced-electron-mass}
\mu_N=\frac{M_1M_2}{M_N}\,,\qquad
{{\mu }_{e}}=\frac{M_N}{1+M_N}
\end{equation}
are the reduced nuclear and electron masses, respectively, and ${\bf
d}$ is the electric dipole moment with respect to the molecular
center of mass,
\begin{equation}
\label{def-d}
\mathbf{d}=-\zeta_e\sum\limits_{i=1}^{L}{{{\mathbf{s}}_{i}}}+{\zeta
_{N}{{\mathbf{S}}}}\,.
\end{equation}

Here we introduced the notations for the reduced electron charge
$\zeta_e$ and reduced nuclear charge $\zeta_N$:
\begin{eqnarray}
\label{def-zeta}
 {{\zeta }_{e}}&=&\frac{ {{M}_{N}}+{{Z}_{N}} }{M_T}\,,\\
 \zeta_N&=&\frac{M_2Z_1-M_1Z_2}{M_N}\,.
\end{eqnarray}

Note that the Hamiltonian\ \eqref{Diatomic_Transformed_Hamiltonian}
contains the potential $V_0$ defined in Eq.\ \eqref{1c}. Here it is assumed that this potential
is expressed in terms of the new variables $\bf S$ and ${\bf
s}_i$.

We point out that this section presents no new results. It only describes the standard change of variables in the molecular Hamiltonian\ \eqref{Diatomic_Hamiltonian} which allows one to
isolate the dynamics of the center of mass of the molecule from the relative dynamics. 




\subsection{Screening of a static external electric field}

Let us consider the screening of
a static homogeneous electric field
$\mathbf{E}_{\rm ext}=\mathbf{E}_0$ at the position ${\bf R}_I$ of the
$I^{\rm th}$ nucleus. The operator of the electric field induced by the other
nucleus and the electrons reads
\begin{equation}\label{Diatomic_E'_2}
\mathbf{E}'_I=-\frac1{Z_I}\nabla_{\mathbf{R}_I} V_0
 =-\frac i{Z_I} \left[ \mathbf{P}_I,H_0 \right]\,,
\end{equation}
where the Hamiltonian $H_0$ is given in Eq.\ \eqref{Diatomic_Transformed_Hamiltonian}.

Note that the Hamiltonian $H_0$ defined in
Eq.~(\ref{Diatomic_Transformed_Hamiltonian}) is independent from
the center-of-mass coordinate ${\bf S}_T$. As a consequence, $[{\bf Q}_T,
H_0]=0$, and Eq.\ \eqref{Diatomic_E'_2} may
be put in the form
\begin{equation}
\label{11}
\mathbf{E}'_I=-\frac i{Z_I}\left[\boldsymbol{\Pi}_I,H_0\right]\,,
\end{equation}
where $\boldsymbol{\Pi}_I$ denotes the truncated momentum operator
\begin{equation}\label{W}
\boldsymbol{\Pi}_I \equiv {\bf P}_I - \frac{M_I}{M_T} {\bf Q}_T
=-(-1)^I{\bf Q} - \frac{M_I}{M_N}\sum_{i=1}^L {\bf q}_i\,.
\end{equation}

Using Eqs.\ \eqref{def-d} and\ \eqref{def-zeta}, one can prove the following commutator relation between $\boldsymbol{\Pi}_I$ and the potential $V_{\rm
rel}$ given in Eq.\ \eqref{6c}
\begin{equation}\label{Diatomic_Commutator_with_V}
\left( 1- \frac{ M_I Z_T }{ M_T Z_I} \right)
\mathbf{E}_0
=-\frac i{Z_I}
 \left[\boldsymbol{\Pi}_I,-{\bf d}\cdot {\bf E}_0
 \right] \,.
\end{equation}

The composition of Eqs.\ \eqref{11} and
\eqref{Diatomic_Commutator_with_V} yields
\begin{equation}\label{Diatomic_Commutator_with_Hrel}
\mathbf{E}'_I+\left( 1- \frac{ M_I Z_T }{ M_T Z_I} \right)\mathbf{E}_0
=-\frac i{Z_I}
 \left[\boldsymbol{\Pi}_I,H_{\rm rel} \right]\,.
\end{equation}

Let $\psi$ be a stationary-state wavefunction describing the molecule in the
center-of-mass frame, namely
\begin{equation}
H_{\rm rel} \psi = {\cal E} \psi\,.
\end{equation}
The expectation value of the commutator on the right-hand side
of Eq.\ \eqref{Diatomic_Commutator_with_Hrel} with respect to $\psi$ vanishes. This
allows us to find the expectation value of the operator $\mathbf{E}'_I$ on
the left-hand side of Eq.\ \eqref{Diatomic_Commutator_with_Hrel},
$\langle \mathbf{E}'_I \rangle \equiv \langle \psi |\mathbf{E}'_I |\psi
\rangle$, so the total electric field at $I^{\rm th}$ nucleus
$\langle \mathbf{E}_I \rangle$ is
%
\begin{equation}\label{Diatomic_Static_Result}
\langle \mathbf{E}_I \rangle \equiv \left\langle \mathbf{E}'_I \right\rangle
+{\mathbf{E}}_{0}=\frac{{{M}_{I}}{{Z}_{T}}}{{{M}_{T}}{{Z}_{I}}}{\mathbf{E}_0}\,.
\end{equation}

This result, derived in a fully quantum mechanical way, agrees with
that obtained in Ref.\ \cite{FlambaumMolecule} with the use of the Ehrenfest theorem.
Note that for a neutral molecule, $Z_T=0$ so the nuclei are completely
screened from the static external field, $\langle \mathbf{E}_I \rangle=0$.
In deriving this result we explicitly assumed that the nuclei
are point-like particles. For real molecules, this screening is
incomplete due to the finite-size effects of the nuclei which are
accounted for by the Schiff moment operator\ \cite{Sandars1,Sandars2,SFK1984,FLAMBAUM1985,FLAMBAUM1986}. In
this paper, however, we do not consider Schiff moment corrections.

\subsection{Off-resonance screening of an oscillating external electric field}
\label{SectIIc}

We now consider the case of an oscillating external electric field
with frequency $\omega$:
\begin{equation}
\label{Eext}
\mathbf{E}_{\rm ext}=\mathbf{E}_0\cos\omega t\,.
\end{equation}
When this field is sufficiently weak, the
time-dependent perturbation theory may be applied to the
Hamiltonian\ \eqref{Diatomic_Transformed_Hamiltonian} with
\begin{equation}\label{18}
V_{\rm rel}(t) = -{\bf d}\cdot {\bf E}_0
\cos\omega t
\end{equation}
treated as the perturbation.

Let $|n\rangle$ be a
complete set of eigenstates of the unperturbed Hamiltonian $H_0$, namely
\begin{equation}
H_0 | n\rangle = {\cal E}_n |n\rangle\,.
\end{equation}

Up to the first order in perturbation theory, the evolution of
the ground state $|0\rangle$ is described by the wavefunction
\begin{equation}\label{20}
\begin{aligned}
\psi(t) &= e^{-i{\cal E}_0 t} \bigg[
|0\rangle \\
& -i \sum_n \int^t d\tau\, e^{-i\omega_{n0}(t-\tau)} |n\rangle \langle n | V_{\rm
rel}(\tau) | 0\rangle
\bigg]\,,
\end{aligned}
\end{equation}
where $\omega_{n0}={\cal E}_n - {\cal E}_0$. In this formula, we
assume that the frequency of the external field $\omega$ is not in
resonance with any transition with energy $\omega_{n0}$. In this case,
it is safe to discard the widths of these states. The resonant case will
be addressed in the next subsection.

Substituting the potential\ \eqref{18} into
Eq.\ \eqref{20}, we find the
expectation value of the operator ${\bf E}'_I$ describing the
induced electric field at the $I^{\rm th}$ nucleus due to
the other nucleus and the electrons,
\begin{equation}\label{21}
\begin{aligned}
\langle {\bf E}'_I \rangle &\equiv
\langle \psi(t)|{\bf E}'_I |\psi(t)\rangle= 2\cos\omega t\\
&\times\sum_n\frac{\omega_{n0}{\rm Re}\langle 0 |{\bf E}'_I |n\rangle
\langle n|{\bf d}\cdot {\bf E}_0|0\rangle}{\omega_{n0}^2 - \omega^2}
\,.
\end{aligned}
\end{equation}

Making use of the identity\ \eqref{11}, one may cast Eq.\ \eqref{21}
in the form
\begin{equation}\label{22}
\langle {\bf E}'_I \rangle =\frac{2\cos\omega t}{Z_I}
\sum_n\frac{\omega_{n0}^2{\rm Im}\left[
\langle 0 |\boldsymbol{\Pi}_I | n\rangle \langle n |{\bf d}\cdot {\bf
E}_0|0\rangle\right]}{\omega_{n0}^2 - \omega^2}
 \,,
\end{equation}
where the operator $\boldsymbol{\Pi}_I$ is defined as in Eq.\ \eqref{W}.
%

Using the identity $\frac{\omega_{n0}^2}{\omega_{n0}^2 - \omega^2}
 = 1+ \frac{\omega^2}{\omega_{n0}^2 - \omega^2}$, the completeness of
the set of states $|n\rangle$ and the relation\ \eqref{Diatomic_Commutator_with_Hrel}, we find the total electric
field at the position of $I^{\rm th}$ nucleus
\begin{equation}\label{E_I}
\begin{aligned}
\langle {\bf E}_I\rangle &\equiv \langle {\bf E}'_I\rangle + {\bf E}_0\cos\omega
t = \frac{M_I Z_T}{M_T Z_I} {\bf E}_0\cos\omega t\\
&+\frac{2\omega^2 \cos\omega t}{Z_I}
 \sum_n \frac{{\rm Im}\, \left[ \langle 0 |\boldsymbol{\Pi}_I | n\rangle \langle n |{\bf d}\cdot {\bf
E}_0|0\rangle\right]}{\omega_{n0}^2 - \omega^2}\,.
\end{aligned}
\end{equation}

We point out that Eq.\ (\ref{E_I}) is valid not only for the Schr\"odinger Hamiltonian (\ref{6b}), but also for the case when the kinetic term for electrons is described by the Dirac Hamiltonian. In the latter case, the relativistic corrections due to the Dirac equation are included in the energies $\omega_{n0}$ and states $|n\rangle$.

Using the explicit form of the Hamiltonian $H_0$ given in Eq.\ \eqref{Diatomic_Transformed_Hamiltonian}, one may prove
the following identity for the operator\ \eqref{W}:
\begin{equation}
\label{W-id}
\boldsymbol{\Pi}_I=\frac{i{{M}_{I}}}{{{\zeta }_{e}}{{M}_{T}}}\left[
 -\mathbf{d}+{\cal M}_I \mathbf{S},H_0 \right]\,,
\end{equation}
where
\begin{equation}\label{calM}
{\cal M}_I =(-1)^I\left( M_N - M_I + Z_N - Z_I\right)\,,
\end{equation}
and the quantities $\bf d$ and $\zeta_e$ are given in Eqs.\
\eqref{def-d} and\ \eqref{def-zeta}, respectively. 

The identity\ \eqref{W-id} allows us to represent the formula\ \eqref{E_I} for the resulting
electric field at the $I^{\rm th}$ as
\begin{equation}\label{Diatomic_Final_Result}
\left\langle {{\mathbf{E}}_I} \right\rangle =
\left[ \frac{M_I Z_T}{M_T Z_I}-\frac{{{\omega }^{2}}{{M}_{I}}}{{\zeta }_{e}M_TZ_I}
 \left(\overset{\scriptscriptstyle\leftrightarrow}{\boldsymbol{\alpha }} -\overset{\scriptscriptstyle\leftrightarrow}{\boldsymbol{\beta }}_I\right)\right]{{\mathbf{E}}_{0}}\cos \omega t\,,
\end{equation}
where
\begin{equation}\label{Diatomic_Polarizability}
\overset{\scriptscriptstyle\leftrightarrow}{\boldsymbol{\alpha
}}=2\sum\limits_{n} \frac{\omega_{n0}}{\omega
_{n0}^{2}-\omega^2}\langle 0|\mathbf{d}|n\rangle
\langle n|\mathbf{d}|0\rangle
\end{equation}
is the molecule's polarizability tensor and
\begin{equation}\label{Diatomic_Beta}
\overset{\scriptscriptstyle\leftrightarrow}{\boldsymbol{\beta
}}_I= 2{\cal M}_I\sum\limits_{n}\frac{\omega_{n0}}{\omega_{n0}^{2}
-\omega^2}\langle0|\mathbf{S}|n\rangle
\langle n|\mathbf{d}|0\rangle\,.
\end{equation}

Eq.\ \eqref{Diatomic_Final_Result} describes the screening of an oscillating
electric field inside a diatomic molecule in the case when the
frequency of the field is off resonance from any molecular
transition. Let us discuss the terms in this formula.

The first term in the brackets has the same form as for the screening of static electric field\ \eqref{Diatomic_Static_Result}. This term is proportional to the
total charge of the molecule $Z_T$ which is vanishing for neutral
molecules.

The second term in the brackets
is specified by the molecular polarizability tensor
$\overset{\scriptscriptstyle\leftrightarrow}{\boldsymbol{\alpha
}}$. This term has the same structure as that in the formula for screening
of electric fields in atoms derived in Ref.\ \cite{FlambaumOscillating}.

The last term in Eq.\  \eqref{Diatomic_Final_Result} is described
by the tensor $\overset{\scriptscriptstyle\leftrightarrow}{\boldsymbol{\beta
}}_I$ defined in Eq.\ \eqref{Diatomic_Beta}. This tensor depends on
the matrix element of the inter-nuclear distance operator ${\bf S}$
which has no analogy in the atomic case. As we will show in the
next section, this term plays a significant role in the screening of
the external field in molecules. In fact, due to the large ratio of nuclear and electron mass $|{\cal M}_I|\approx \left(M_N-M_I \right)/m_e \gg 1$, $\overset{\scriptscriptstyle\leftrightarrow}{\boldsymbol{\beta
}}_I$ usually dominates over $\overset{\scriptscriptstyle\leftrightarrow}{\boldsymbol{\alpha
}}$. As will be discussed in Sect.\ \ref{Screening of the external electric field in different frequency regimes}, if the external field's frequency is in the rotational or vibrational regimes ($10^{-5} - 10^{-3}\, {\rm a.u.}$) then $|\overset{\scriptscriptstyle\leftrightarrow}{\boldsymbol{\beta
}}_I| \gg \overset{\scriptscriptstyle\leftrightarrow}{\boldsymbol{\alpha
}}$. Only the case where the field's frequency is in the electronic transition regime ($\sim 0.1\,{\rm a.u.}$) that $\overset{\scriptscriptstyle\leftrightarrow}{\boldsymbol{\beta
}}_I$ and $\overset{\scriptscriptstyle\leftrightarrow}{\boldsymbol{\alpha
}}$ become comparable.

In conclusion of this section, we rewrite our results\ \eqref{Diatomic_Final_Result}--\eqref{Diatomic_Beta} with the
fundamental constants $\hbar$, $e$ and $m_e$ explicitly shown
\begin{subequations}\label{full}
\begin{eqnarray}
\left\langle {{\mathbf{E}}_I} \right\rangle &=& \left[ \frac{{{M}_I}{{Z}_{T}}}{M_TZ_I}
-\frac{{{\omega }^{2}}m_e{{M}_I}}{e^2{\zeta }_{e}M_TZ_I}
 \left(\overset{\scriptscriptstyle\leftrightarrow}{\boldsymbol{\alpha }} -\overset{\scriptscriptstyle\leftrightarrow}{\boldsymbol{\beta }}_I\right)\right]{{\mathbf{E}}_{\rm ext}}
\,,\label{E_I_full}
\\
\overset{\scriptscriptstyle\leftrightarrow}{\boldsymbol{\alpha
}}&=&\frac{2}{\hbar}\sum\limits_{n}
\frac{{{\omega }_{n0}}}{\omega
_{n0}^{2}-{{\omega }^{2}}} \langle 0|\mathbf{d}|n\rangle
\langle n|\mathbf{d}|0\rangle \,,
\label{alpha}
\\
\overset{\scriptscriptstyle\leftrightarrow}{\boldsymbol{\beta
}}_I &=& \frac{2{\cal M}_I}{\hbar} 
 \sum\limits_{n}
\frac{{{\omega }_{n0}}\langle 0| e\mathbf{S}|n\rangle
\langle n|\mathbf{d}|0\rangle}{\omega _{n0}^{2}-{{\omega }^{2}}}
\,,
\label{beta}
\end{eqnarray}
\end{subequations}
where ${\cal M}_I =(-1)^I\left[ m_e^{-1}(M_N-M_I)+Z_N-Z_I\right]$,
$\zeta_e=M_T^{-1}\left(M_N+m_eZ_N\right)$ and
$\mathbf{d}=-\zeta_ee\sum\limits_{i=1}^{L}{{{\mathbf{s}}_{i}}}+{\zeta _{N}e{{\mathbf{S}}}}$.

Since $m_e\ll M_I$, one can make the (good) approximations $\zeta_e\approx 1$, ${\cal M}_1 \approx -M_2/m_e$, ${\cal M}_2\approx M_1/m_e$ and $M_T\approx M_N$. Note that, with these approximations, the final term in the bracket in Eq.\ \eqref{E_I_full} may be written as
\begin{equation}
\begin{aligned}
\frac{{{\omega }^{2}}m_e{{M}_I}}{e^2{\zeta }_{e}M_TZ_I}\overset{\scriptscriptstyle\leftrightarrow}{\boldsymbol{\beta
}}_I&\approx\frac{2(-1)^I\omega^2\mu_N}{e^2\hbar Z_I}\\
&\times\sum\limits_{n}
\frac{{{\omega }_{n0}}\langle 0| e\mathbf{S}|n\rangle
\langle n|\mathbf{d}|0\rangle}{\omega _{n0}^{2}-{{\omega }^{2}}}\,,
\end{aligned}
\end{equation}
where $\mu_N$ is the reduced nuclear mass defined in Eq.\ \eqref{reduced-electron-mass}.


\subsection{Resonance enhancement of an oscillating external electric field}
\label{Sect2D}

When the frequency of the external electric field\ \eqref{Eext}
approaches one of the molecular transition frequencies, $\omega = \omega_{n0}$, the width of this state $\Gamma_n$ cannot be ignored in perturbative calculations. In this case, the
wavefunction\ \eqref{20} should be modified to read
\begin{equation}\label{psi+Gamma}
\begin{aligned}
\psi(t) &= e^{-i{\cal E}_0 t} \bigg[
|0\rangle -i \sum_k \int^t \bigl(d\tau\\
&  \times e^{-i(\omega_{k0}-i\Gamma_k/2)(t-\tau)}|k\rangle \langle k | V_{\rm
rel}(\tau) | 0\rangle\bigr)
\bigg]\,.
\end{aligned}
\end{equation}

Using the explicit form of the potential\ \eqref{18}, we find the
expectation value of the operator ${\bf E}'_I$ describing the
electric field at the $I^{\rm th}$ nucleus induced by the electrons and the
other nucleus
\begin{equation}\label{31}
\begin{aligned}
\langle {\bf E}'_I \rangle &=
\sum_k[-\cos(\omega t)  f_k(\omega) +\frac{\Gamma_k}2 \sin(\omega
t)g_k(\omega) ]\\
&\times  {\rm Re}
\langle 0| {\bf E}'_I | k \rangle \langle k| -{\bf d}\cdot{\bf
E}_0 |0\rangle\,,
\end{aligned}
\end{equation}
where
\begin{subequations}
\begin{eqnarray}
f_k(\omega) &=& \frac{\omega_{k0}+\omega}{(\omega_{k0}+\omega)^2 -({\Gamma_k}/{2})^2}
\nonumber\\
&&+\frac{\omega_{k0}-\omega}{(\omega_{k0}-\omega)^2
-({\Gamma_k}/{2})^2}\,,\\
g_k(\omega) &=& \frac{1}{(\omega_{k0}+\omega)^2 -({\Gamma_k}/{2})^2}\nonumber\\
&&-\frac{1}{(\omega_{k0}-\omega)^2 -({\Gamma_k}/{2})^2}\,.
\label{fg}
\end{eqnarray}
\end{subequations}
Note that in Eq.\ \eqref{31}, we keep only the terms which are not
suppressed by the small factor $e^{-\Gamma_k t/2}$.

It is natural to assume that all linewidths $\Gamma_k$ are much
smaller than the corresponding energies, $\Gamma_k\ll
\omega_{k0}$. Under this assumption, the leading term in the
functions $f_k\left(\omega\right)$ and $g_k\left(\omega\right)$ are
\begin{subequations}
\begin{eqnarray}
f_k(\omega) &=& \left\{
\begin{array}{ll}
1/2\omega\,,\qquad& k = n\\
{2\omega_{k0}}/{\left(\omega_{k0}^2 - \omega^2\right)}\,, & k\ne n
\end{array}\,,
 \right.
\\
g_k(\omega) &=& \left\{
\begin{array}{ll}
-4/{\Gamma_n^2}\,,\qquad& k = n\\
-{4\omega_{k0}\omega}/{(\omega_{k0}^2 - \omega^2)^2}\,, & k\ne
n
\end{array}\,.
 \right.
\end{eqnarray}
\end{subequations}

Substituting these functions into Eq.\ \eqref{31} and using the
identity\ \eqref{11}, we find the total field at the $I^{\rm th}$ nucleus
$\langle {\bf E}_I \rangle \equiv {\bf E}_0\cos(\omega t) + \langle {\bf E}'_I
\rangle$,
\begin{equation}\label{35}
\begin{aligned}
\langle {\bf E}_I \rangle &=\frac{M_I Z_T}{M_T Z_I} {\bf E}_0
\cos\omega t \\
&+\frac{2\omega^2 \cos\omega t}{Z_I}  \sum_{k\ne n}
\frac{{\rm Im}\left[\langle 0|\boldsymbol{\Pi}_I |k\rangle \langle
k |{\bf d}\cdot {\bf E}_0|0\rangle\right]}{\omega_{k0}^2 - \omega^2}
\\
&-\frac{3\cos\omega t}{2Z_I} {\rm Im}\left[\langle 0|\boldsymbol{\Pi}_I |n\rangle \langle
n |{\bf d}\cdot {\bf E}_0|0\rangle\right]
\\
&-\frac{2\omega\sin\omega t}{Z_I\Gamma_n} 
{\rm Im}\left[ \langle 0|\boldsymbol{\Pi}_I |n\rangle \langle
n |{\bf d}\cdot {\bf E}_0|0\rangle\right]\,.
\end{aligned}
\end{equation}

In deriving this result, we have employed the identity\ \eqref{Diatomic_Commutator_with_V} and taken into account the
completeness of the set of states $|k\rangle$.

Different terms in the expression\ \eqref{35} play different role
in the screening or resonant enhancement of the electric field at
the $I^{\rm th}$ nucleus in the molecule. Let us discuss them
separately.

The term in the first line of Eq.\ \eqref{35} is present only for charged molecules with $Z_T\ne 0$. In the rest of this
subsection we consider neutral molecules for
which this term vanishes.

The term in the second line is responsible
for the screening of the external field due to the states which
are off resonance with frequency $\omega$. Strictly speaking, this term cannot be expressed via the tensors\ \eqref{alpha} and\ \eqref{beta} since the sum does not contain the state $|n\rangle$, which is on resonance with the external field.

The terms in the last two lines arise from this
state $|n\rangle$ which is in resonance with the external
field. Between these terms, the last one dominates because the width
is typically much smaller than the energy, $\Gamma_n\ll \omega$.
Moreover, if $|n\rangle$ is a rotational or vibrational state then $\Gamma_n$ is typically very small and the factor $\omega/\Gamma_n$
in the last term of Eq.\ \eqref{35} makes it to dominate over
all other terms. We can thus write (with the constant $e$ and $m_e$ restored)
\begin{equation}\label{36}
\langle {\bf E}_I \rangle \approx
\frac{\omega^2m_eM_I}{\zeta_ee^2 Z_I M_T \Gamma_n}
\overset{\scriptscriptstyle\leftrightarrow}{\boldsymbol{\gamma }}_I
{\bf E}_0 \sin\omega t\,,
%
\end{equation}
where we have introduced the tensor
\begin{equation}
\label{gamma-tensor}
\overset{\scriptscriptstyle\leftrightarrow}{\boldsymbol{\gamma
}}_I = 2\langle 0| \mathbf{d} |n\rangle \langle n| \mathbf{d} |0\rangle 
-2{\cal M}_I \langle 0|\mathbf{S}|n\rangle \langle
n |{\bf d}|0\rangle\,.
\end{equation}
In deriving Eq.\ \eqref{36} we applied the identity\ \eqref{W-id}.


The electric field at the $I^{\rm th}$ nucleus\ \eqref{36} has the following features: (i) The external field may
be enhanced by many orders of magnitude due to the resonance factor
$\omega/\Gamma$. (ii) The phase of the resulting field is shifted by
$\pi/2$ with respect to the applied field. This is a typical
resonance phase shift which occurs in damped driven oscillations.
Both these features are already known for an atom in an oscillating
electric field \cite{SamsonovResonant}. Here we establish similar results for molecules.

We note that although in the resonance case, the external field may be significantly enhanced thanks to the smallness of the width $\Gamma_n$, it certainly does not mean that one can induce an arbitrarily large field at a nucleus by using an arbitrarily strong external field. This is because for strong fields, perturbation theory, in the framework of which the results of this section were derived, breaks down. The condition for the applicability of perturbation theory reads 
\begin{equation}
\label{contraint-omega}
    \Omega \ll \Gamma_n\,,
\end{equation}
where $\Omega \equiv |\langle 0|\mathbf{d}\cdot \mathbf{E}_0|n\rangle|$ is the Rabi frequency (see Ref.\ \cite{SamsonovResonant} for more detail).

If the condition $\Omega \ll \Gamma_n$ is not met then $\langle\mathbf{E}_I\rangle$ may be calculated non-perturbatively by considering the states $|0\rangle$ and $|n\rangle$ as forming a two-level quantum system. The result is
\begin{equation}
    \label{With_Omega_Rabi}
\langle {\bf E}_I \rangle \approx
\frac{\omega^2m_eM_I}{\zeta_ee^2 Z_I M_T }\frac{\Gamma_n}{\Gamma_n^2+2\Omega^2}
\overset{\scriptscriptstyle\leftrightarrow}{\boldsymbol{\gamma }}_I
{\bf E}_0 \sin\omega t\,.
\end{equation}

We note that for the weak external field $E_0$ such that $\Omega\ll\Gamma_n$, Eq. \ (\ref{With_Omega_Rabi}) reduces to Eq.\ (\ref{36}) derived perturbatively. For a strong external field such that $\Omega\gg\Gamma_n$ the resulting electric field at nucleus $\langle\mathbf{E}_I\rangle$ becomes inversely proportional to the applied field $E_0$.

According to Eq.\ \eqref{With_Omega_Rabi}, the field $\langle\mathbf{E_I}\rangle$, considered as a function of $E_0$ reaches its maximum at 
\begin{equation}\label{E_crit}
    E_0=E_{\rm crit}\equiv\frac{\Gamma_n}{\sqrt{2}d}\,,
\end{equation}
where $d=|\langle 0|\mathbf{d}\cdot \hat{\mathbf{k}}|n\rangle|$. Here, $\hat{\mathbf{k}}$ is the unit vector in the direction of $\mathbf{E}_0$. The maximal value of $E_I\equiv|\langle\mathbf{E_I}\rangle|$, corresponding to $E_0=E_{\rm crit}$ is
\begin{equation}\label{E_I_max}
    E_I^{\rm max}\approx
\frac{\omega^2m_eM_I}{\sqrt{8}\zeta_ee^2 Z_I M_T d}
|\overset{\scriptscriptstyle\leftrightarrow}{\boldsymbol{\gamma }}_I
\hat{\bf k}|\sin\omega t\,.
\end{equation}
It is important to note that this field is independent of the width of the excited state $\Gamma_n$.
%

\section{Numerical estimates for diatomic molecules}\label{Diatomic_Numerical}

In this section, the screening of an electric field in some diatomic molecules will be considered. To find the total electric field at the $I^{\rm th}$ nucleus (\ref{Diatomic_Final_Result}), one needs to evaluate the tensors $\overset{\scriptscriptstyle\leftrightarrow}{\boldsymbol{\alpha }}$, $\overset{\scriptscriptstyle\leftrightarrow}{\boldsymbol{\beta }}_I$ and $\overset{\scriptscriptstyle\leftrightarrow}{\boldsymbol{\gamma }}_I$
given in Eqs.\ \eqref{Diatomic_Polarizability},\ \eqref{Diatomic_Beta} and\ \eqref{gamma-tensor}.
In general, to accurately compute these tensors one needs to apply sophisticated numerical methods. Our goal is, however, to give crude semi-analytic estimates of some leading contributions to these tensors. For this purpose, we first develop a representation of these tensors in the Born-Oppenheimer approximation and then present numerical estimates for some simple molecules.

\subsection{Molecular polarizability in the Born-Oppenheimer approximation}

The leading contributions to the tensors\ \eqref{Diatomic_Polarizability},\ \eqref{Diatomic_Beta} and \ \eqref{gamma-tensor} may be estimated
in the Born-Oppenheimer approximation. In this approximation, the motion of the molecule in any state $|n\rangle$ may be separated into rotational, vibrational and electronic modes. The rotational motion is described by
the Wigner $D$-matrix $D^{J_n}_{M_n\Lambda_n}(\Theta)$ depending
on the set of Euler angles $\Theta$ which describes the molecule's orientation with respect to the fixed laboratory frame ($J_n$ is the molecule's total angular momentum
quantum number, $M_n$ is the projection of this angular momentum onto the
laboratory $z$-axis and $\Lambda_n$ is the projection of the electronic angular momentum onto the inter-nucler axis)
whereas the vibrational and electronic motions may be described by the ket
$|\mu_n\Lambda_n\nu_n\rangle$ ($\nu_n$ is the vibrational quantum number which, generally,
depends on $J_n$, $\nu_n=\nu_n(J_n)$, and $\mu_n$ denotes all other quantum
numbers). We write
\begin{equation}
\label{psi-BO}
|n\rangle=\sqrt{\frac{2J_n+1}{8\pi^2}}D^{J_n}_{M_n\Lambda_n}(\Theta)|\mu_n\Lambda_n\nu_n\rangle\,.
\end{equation}
where the coefficient $\sqrt{\left(2J_n+1\right)/8\pi^2}$ is the normalization constant for the Wigner $D$-matrix. The ket $|\mu_n\Lambda_n\nu_n\rangle$ is assumed to be properly normalized.

The ket $|\mu_n\Lambda_n\nu_n\rangle$ may be considered as a
wavefunction of the molecule in the rotating frame. This
wavefunction factorizes into the
vibrational $\psi^{\mu_n \Lambda_n}_{\nu_n}(S)$ and electronic
$\phi_{\mu_n\Lambda_n}(S,\mathbf{s}_i)$ parts,
\begin{equation}
|\mu_n\Lambda_n\nu_n\rangle=S^{-1}\psi^{\mu_n
\Lambda_n}_{\nu_n}(S)\phi_{\mu_n\Lambda_n}(S,\mathbf{s}_i)\,.
\end{equation}

In this representation, we ignore all spins of electrons and
nuclei. This approximation usually gives acceptable accuracy for low rotational
levels  \cite{BRIEGER}.

In literature, the polarizablity tensor $\overset{\scriptscriptstyle\leftrightarrow}{\boldsymbol{\alpha }}$ is often calculated in the frame rotating with the molecule (this frame is defined with respect to the laboratory by the angular coordinates $\Theta$). However, the external electric field is naturally defined with respect to the laboratory frame. Thus, one needs to express the laboratory-frame components of the molecular polarizability tensor
via its rotating-frame components. 


The spherical components of the vector $\mathbf{d}$ in the laboratory ($l$) and rotating ($r$) frames will be denoted by $d^{(l)}_q$ and $d^{(r)}_q$ ($q=0,\pm 1$), respectively.
They are related to each other by the formula \cite{BRIEGER}

\begin{equation}d_{p}^{(l)}=\sum\limits_{q}{{{\left( -1 \right)}^{p-q}}D_{pq}^1( \Theta  )d_{q}^{(r)}}\,.
\end{equation}

This formula, together with Eq.\ (\ref{psi-BO}) and the standard identity for the Wigner $D$-matrices
\cite{mizushima1975theory,edmonds2016angular}
\begin{equation}
\begin{aligned}
  & \frac{1}{8{{\pi }^{2}}}\int{\bar{D}_{{M}'{\Lambda }'}^{{{J}'}}( \Theta )D_{pq}^{j}( \Theta  )D_{M\Lambda }^{J}( \Theta  )d\Theta } \\ 
 & ={{\left( -1 \right)}^{2{J}'-{M}'-{\Lambda }'}}\left( \begin{matrix}
   {{J}'} & j & J  \\
   -{M}' & p & M  \\
\end{matrix} \right)\left( \begin{matrix}
   {{J}'} & j & J  \\
   -{\Lambda }' & q & \Lambda   \\
\end{matrix} \right) \,,
\end{aligned}
\end{equation}
allows us to relate the laboratory-frame components of the tensor $\langle0|\mathbf{d}|n\rangle\langle n|\mathbf{d}|0\rangle$ 
to its rotating-frame components
\begin{equation}\label{42}
\begin{aligned}
  &\langle0|d_{p}^{(l)}|n\rangle \langle n|\bar{d}_{q}^{(l)}|0\rangle
   ={{\delta }_{pq}}\left( 2{{J}_{0}}+1 \right)\left( 2{{J}_{n}}+1 \right) \\
  &\times {{\left( \begin{matrix}
   {{J}_{0}} & 1 & {{J}_{n}}  \\
   -{{M}_{0}} & p & {{M}_{n}}  \\
\end{matrix} \right)}^{2}}{{\left( \begin{matrix}
   {{J}_{0}} & 1 & {{J}_{n}}  \\
   -{{\Lambda }_{0}} & {{\Lambda }_{0}}-{{\Lambda }_{n}} & {{\Lambda }_{n}}  \\
\end{matrix} \right)}^{2}} \\
  &\times \left|\langle\mu_0\Lambda_0 \nu_0 | d_{\Lambda_{0}-\Lambda_n}^{(r)}  |{{\mu }_{n}}{\Lambda_{n}}{\nu_n}\rangle\right|^2\,.
\end{aligned}
\end{equation}

If the molecule is non-polarized in the ground state, the expression (\ref{42}) should be averaged over the quantum number $M_0$ and summed over the quantum number $M_n$. 
In this case, the molecular polarizability tensor (\ref{Diatomic_Polarizability}) 
has only the scalar part $\alpha(\omega)$, namely
\begin{equation}
\label{alpha-diagonal}
    \overset{\scriptscriptstyle\leftrightarrow}{\boldsymbol{\alpha }}(\omega)
     = \alpha(\omega) {\bf 1}_{3\times3}\,.
\end{equation}
Physically, this implies that the induced electric field in the non-polarized molecule can only be parallel to the external electric field. 

Without loss of generality, it may be assumed that the external
field $\mathbf{E}_{\rm ext}$ is directed along the laboratory
$z$-axis. For simplicity, it will also be assumed that the ground
state is in the $\Sigma$ ($\Lambda=0$) configuration and has total
angular momentum $J_0=0$ (this assumption forces $M_0=0$ and
$J_n=1$). Under these assumptions, the only relevant component of the tensor (\ref{42}) is
\begin{equation}\label{00_Component}
\begin{aligned}
  &\langle0|d^{(l)}_0|n\rangle\langle n|\bar{d}^{(l)}_0|0\rangle
    =\delta_{0M_n}\left( \begin{matrix}
   0 & 1 & 1  \\
   0 & -{{\Lambda }_{n}} & {{\Lambda }_{n}}  \\
\end{matrix} \right)^2 \\
  &\times \left|\langle{\mu _0\Sigma \nu _0 } |d_{-{{\Lambda }_{n}}}^{(r)}  |{{{\mu }_{n}}{{\Lambda }_{n}}\nu_n}\rangle\right|^{2}\,,
\end{aligned}
\end{equation}
where the vibrational quantum number $\nu_0$ corresponds to the total momentum quantum number $J=0$ whereas $\nu_n$ is attributed to $J=1$.

Substituting Eq.\ \eqref{00_Component} into Eq.\ \eqref{Diatomic_Polarizability}, we obtain
the expression for the polarizability tensor in the laboratory frame via the components of this tensor in the molecular (rotating) frame
\begin{equation}
\label{alpha00}
\alpha(\omega)=\frac13 \left[\alpha^{\parallel }_{\rm el}( \omega  )+2\alpha^{\perp }_{\rm el}( \omega )+\alpha^{\parallel }_{\rm vib}( \omega  )+\alpha^{\parallel }_{\rm rot}( \omega  )\right],
\end{equation}
where $\alpha^{\parallel }_{\rm el}$ and $\alpha^{\perp }_{\rm el}$ are the parallel and perpendicular components with respect to the molecular axis arising due to electronic excitations;
$\alpha^{\parallel }_{\rm vib}$ and $\alpha^{\parallel }_{\rm rot}$ are the terms originating from the vibrational and rotational states. Explicitly, these quantities read
\begin{subequations}\label{45}
\begin{eqnarray}\label{Definitions of Polarizabilities}
  \alpha^{\parallel }_{\rm el}( \omega )&=&\sum\limits_{\begin{smallmatrix}
 {{\mu }_{n}}\ne {{\mu }_{0}} \\
 {{\nu }_{n}}
\end{smallmatrix}}
 \frac{2\omega_{n0}}{\omega _{n0}^{2}-{{\omega }^{2}}}{{\left| \left\langle  {{\mu }_{0}}\Sigma {{\nu }_{0}} \right|d_{0}^{(r)}\left| {{\mu }_{n}}\Sigma {{\nu }_{n}} \right\rangle  \right|}^{2}} 
 \label{45a},~~~~~\\
\alpha^{\perp }_{\rm el}( \omega)&=&\sum\limits_{
 \mu_n,\nu_n }{\frac{2{{\omega }_{n0}}}{\omega _{n0}^{2}-{{\omega }^{2}}}{{\left| \left\langle  {{\mu }_{0}}\Sigma {{\nu }_{0}} \right|d_{-1}^{(r)}\left| {{\mu }_{n}}\Pi {{\nu }_{n}} \right\rangle  \right|}^{2}}} , \label{45b}\\
\alpha^{\parallel }_{\rm vib}( \omega )&=&\sum\limits_{{{\nu }_{n}}\ne {{\nu }_{0}}}{\frac{2{{\omega }_{n0}}}{\omega _{n0}^{2}-{{\omega }^{2}}}{{\left| \left\langle  {{\mu }_{0}}\Sigma {{\nu }_{0}} \right|d_{0}^{(r)}\left| {{\mu }_{0}}\Sigma {{\nu }_{n}} \right\rangle  \right|}^{2}}} ,\label{45c}\\
\alpha^{\parallel }_{\rm rot}( \omega )&=&\frac{2{{\omega }_{\nu_0\nu_0'}}}{\omega _{\nu_0\nu_0'}^{2}-{{\omega }^{2}}}{{\left| \left\langle  {{\mu }_{0}}\Sigma {{\nu }_{0}}\right|d_{0}^{(r)}\left| {{\mu }_{0}}\Sigma {{\nu }'_{0}} \right\rangle  \right|}^{2}} , \label{45d}
\end{eqnarray}
\end{subequations}
where the vibrational quantum numbers $\nu_0$ and $\nu'_0$ correspond to $J=0$ and $J=1$, respectively.

As we will show in the end of this section, the representation for the molecular
polarizablity (\ref{alpha00}) appears useful, as the values of the quantities (\ref{45}) may be either easily calculated or found in literature. Similar representations may be developed for the tensors (\ref{Diatomic_Beta}) and (\ref{gamma-tensor}). Following the same procedure as above, it is possible to prove that each of these tensors is proportional to the unit matrix,
\begin{equation}
    \overset{\scriptscriptstyle\leftrightarrow}{\boldsymbol{\beta}_I}(\omega)
     = \beta_I(\omega){\bf 1}_{3\times 3}\,,\qquad
     \overset{\scriptscriptstyle\leftrightarrow}{\boldsymbol{\gamma}_I}(\omega)
     = \gamma_I(\omega){\bf 1}_{3\times 3}\,.
     \label{48-}
\end{equation}

For the quantity $\beta_I(\omega)$ we find 
\begin{equation}\label{47}
\beta_I(\omega)=\frac{\mathcal{M}_I}3 \left[\beta^{\parallel}_{\rm vib}( \omega)+\beta^{\parallel}_{\rm rot}( \omega  )\right]\,,
\end{equation}
where
\begin{subequations}
\label{47-}
\begin{eqnarray}
  \beta^{\parallel }_{\rm vib}( \omega  )&=&2\sum\limits_{{{\nu }_{n}}\ne {{\nu }_{0}}}{\frac{{{\omega }_{n0}}}{\omega _{n0}^{2}-{{\omega }^{2}}}\left\langle  {{\mu }_{0}}\Sigma {{\nu }_{0}} \right|S\left| {{\mu }_{0}}\Sigma {{\nu }_{n}} \right\rangle } \nonumber\\
  &\times & \left\langle  {{\mu }_{0}}\Sigma {{\nu }_{n}} \right|d_{0}^{(r)}\left| {{\mu }_{0}}\Sigma {{\nu }_{0}} \right\rangle  \,,\\
  \beta^{\parallel }_{\rm rot}( \omega )&=&\frac{2{\omega }_{\nu_0\nu_0'}}{\omega _{\nu_0\nu_0'}^{2}-{{\omega }^{2}}}\left\langle  {{\mu }_{0}}\Sigma {{\nu }_{0}}\right|S\left| {{\mu }_{0}}\Sigma \nu'_0 \right\rangle \nonumber \\
  &\times &\left\langle  \mu_0\Sigma \nu'_0 \right|d_{0}^{(r)}\left| {{\mu }_{0}}\Sigma {{\nu }_{0}} \right\rangle\,.
\label{47-b}
\end{eqnarray}
\end{subequations}

In contrast to the quantities (\ref{Diatomic_Polarizability}) and (\ref{Diatomic_Beta}), there is no summation over states in the formula of the $\gamma$-tensor (\ref{gamma-tensor}) since only one pair of states is in resonance with the external field. 

It is of interest to consider the external electric field in resonance with the lowest rotational state. In this case, we have
\begin{equation}\label{46-}
\begin{aligned}
\gamma_I(\omega)&=
\frac{2}{3}\left|\left\langle  {{\mu }_{0}}\Sigma {{\nu }_{0}}\right|d_{0}^{(r)}\left| {{\mu }_{0}}\Sigma {{\nu }_{n}}\right\rangle\right|^2\\
&-\frac23\mathcal{M}_I\left\langle  {{\mu }_{0}}\Sigma {{\nu }_{0}}\right|S\left| {{\mu }_{0}}\Sigma {{\nu }_{n}}\right\rangle\\
&\times\left\langle  {{\mu }_{0}}\Sigma {{\nu }_{n}}\right|d_{0}^{(r)}\left| {{\mu }_{0}}\Sigma {{\nu }_{0}} \right\rangle\,.
\end{aligned}
\end{equation}

In this formula, the second term on the right-hand side dominates over the first one since $|{\cal M}_I|\gg1$,
\begin{equation}\label{46}
\begin{aligned}
\gamma_I(\omega)&\approx
-\frac23\mathcal{M}_I\left\langle  {{\mu }_{0}}\Sigma {{\nu }_{0}}\right|S\left| {{\mu }_{0}}\Sigma {{\nu }_{n}}\right\rangle\\
&\times\left\langle  {{\mu }_{0}}\Sigma {{\nu }_{n}}\right|d_{0}^{(r)}\left| {{\mu }_{0}}\Sigma {{\nu }_{0}} \right\rangle\,.
\end{aligned}
\end{equation}

We point out, however, that the first term in Eq.\ (\ref{46-}) cannot be ignored if the external electric field is in resonance with electronic transitions in the molecule. In this case, the second term in Eq.\ (\ref{46-}) vanishes so the sole contribution to $\gamma_I\left(\omega\right)$ comes from its first term.

%

\subsection{Screening of the external electric field in different frequency regimes}\label{Screening of the external electric field in different frequency regimes}

Recall that the molecular spectra have three typical frequency scales:  $\omega_{\rm rot}$ associated with the rotational transitions, $\omega_{\rm vib}$ associated with the vibrational transitions and $\omega_{\rm el}$ associated with the electronic transitions. Normaly, $\omega_{\rm rot}\ll \omega_{\rm vib} \ll \omega_{\rm el}$. It is interesting to consider the screening of external electric fields with frequencies in these three different regimes. 

As was shown in the previous subsection, the induced electric field at the $I^{\rm th}$ nucleus is parallel to the external electric field. The magnitude of this field may be written as
%
%
\begin{equation}\label{SumUp1}
E_I=\sigma_I(\omega)E_0\cos\omega t\,,
\end{equation}
where the suppression factor $\sigma_I$ is defined as
\begin{equation}\label{sigmaI}
\sigma_I(\omega)=\frac{M_IZ_T}{M_TZ_I} -\frac{\omega^2M_I}{\zeta_eM_TZ_I}\left[\alpha(\omega)-\beta_I(\omega)\right]\,.
\end{equation}
The first term here is $\omega$-independent and is non-vanishing only for charged molecules. Although this term is important in the total formula for the suppression factor, it is more interesting to analyze the other terms in Eq.\ (\ref{sigmaI}) which are non-vanishing both for charged and neutral molecules and are $\omega$-dependent. Therefore, to the end of this subsection we will restrict ourselves to the case $Z_T=0$. We will also work in the limit $m_e\ll M_I$ in which $\zeta_e\approx 1$, $\mathcal{M}_I\approx(-1)^I\left(M_N-M_I\right)$ and $M_T\approx M_N$.

\subsubsection{Screening of rotational-range-frequency fields}

Let us consider an external electric field with frequency of order of the lowest rotational transition frequencies, $\omega\sim\omega_{\nu_0\nu'_0}$. In this case, because the factor $\omega_{n0}/\left(\omega_{n0}^2-\omega^2\right)$ in Eq.\ (\ref{Diatomic_Polarizability}) scales as energy inverse, 
the dominant contribution to $\alpha(\omega)$ comes from Eq.\ \eqref{45d} and the dominant contribution to $\beta_I(\omega)$ comes from Eq.\ (\ref{47-b}). Since the matrix elements $\left\langle  {{\mu }_{0}}\Sigma {{\nu }_{0}}\right|S\left| {{\mu }_{0}}\Sigma \nu'_0 \right\rangle$ and $\left\langle  \mu_0\Sigma \nu_0\right|d^{(r)}_0\left| \mu_0\Sigma \nu'_0\right\rangle$ are of the same order, and since $|\mathcal{M}_I|\gg 1$, the dominant contribution to the suppression factor (\ref{sigmaI}) comes from $\beta_I(\omega)$. Thus, the leading contribution to the suppression factor (\ref{sigmaI}) is given by the term (\ref{47-b}), namely
\begin{equation}
\label{sigma-I}
\sigma_I^{\rm rot}\left(\omega\right)\approx \left(-1\right)^I\frac{2\omega^2\mu_N}{3Z_I}\frac{\bar{\omega}\bar{S}\bar d }{\bar{\omega}^2-\omega^2}\,,
\end{equation}
where 
\begin{equation}\label{56}
\bar{\omega}\equiv\omega_{\nu_0\nu'_0}=\frac{1}{\mu_N\bar{S}^2}
\end{equation}
is the rotational energy with $J=1$, 
\begin{equation}\label{57}
\bar S=\left\langle  {{\mu }_{0}}\Sigma {{\nu }_{0}}\right|S\left| {{\mu }_{0}}\Sigma {{\nu }_{0}}\right\rangle
\end{equation}
is the ground state mean inter-nuclear distance and 
\begin{equation}\label{58}
\bar d=\left\langle  {{\mu }_{0}}\Sigma {{\nu }_{0}}\right|d^{(r)}_0\left| {{\mu }_{0}}\Sigma {{\nu }_{0}}\right\rangle
\end{equation}is the ground state mean electric dipole (in the direction of $\mathbf{S}$). Note that we have invoked the rigid-rotor approximation which allows us to replace $\nu_0'$ by $\nu_0$ in the definition of $\bar S$ and $\bar d$. We have also made the approximation $M_I{\cal M}_I/M_T\approx(-1)^I\mu_N$.

\subsubsection{Screening of vibrational-range-frequency fields} 

In the case where $\omega$ is of the order of a vibrational energy, $\omega\sim\omega_{\rm vib}$, contributions to $\alpha(\omega)$ from $\alpha^{\parallel}_{\rm el}$ and $\alpha^{\perp}_{\rm el}$ are negligible whereas the terms $\alpha^{\parallel}_{\rm vib}$ and $\alpha^{\parallel}_{\rm rot}$ are significant. 
In the expression (\ref{47}) for $\beta_I$, both terms $\beta^{\parallel}_{\rm vib}$ and $\beta^{\parallel}_{\rm rot}$ contribute. Again, since $|\mathcal{M}_I|\gg 1$, $\beta_I$ dominates over $\alpha$. Thus, in this case, the leading contributions to the suppression factor (\ref{sigmaI}) are
\begin{subequations}\label{Vibrational Sigma 2}
\begin{eqnarray}
\sigma_I(\omega)&=&\sigma_I^{\rm vib}(\omega)+\sigma_I^{\rm rot}(\omega)\,,\label{sigma_vib}\\
\sigma_I^{\rm vib}(\omega)&\approx&\left(-1\right)^I\frac{2\omega^2\mu_N}{3Z_I}\sum\limits_{{{\nu }_{n}}\ne {{\nu }_{0}}}\frac{{{\omega }_{n0}}S_{\nu_0}^{\nu_n}\left(d^r_0\right)_{\nu_n}^{\nu_0}}{\omega _{n0}^{2}-{{\omega }^{2}}}\,,~~~~~
\label{60b}
\end{eqnarray}
\end{subequations}
where $\bar\omega$, $\bar S$ and $\bar d$ are given in Eqs.\ (\ref{56}), (\ref{57}) and (\ref{58}), respectively, and the quantities $S_{\nu_0}^{\nu_n}$ and $\left(d^r_0\right)$ are defined by
\begin{equation}\label{60}
    S_{\nu_0}^{\nu_n}=\left\langle  {{\mu }_{0}}\Sigma {{\nu }_{0}}\right|S\left| {{\mu }_{0}}\Sigma {{\nu }_{n}}\right\rangle
\end{equation}
and
\begin{equation}\label{61}
    \left(d^r_0\right)_{\nu_n}^{\nu_0}=\left\langle  {{\mu }_{0}}\Sigma {{\nu }_{n}}\right|d^{(r)}_0\left| {{\mu }_{0}}\Sigma {{\nu }_{0}}\right\rangle\,.
\end{equation}

The quantity $\sigma_I^{\rm rot}(\omega)$ in Eq.\ (\ref{sigma_vib}) is defined in Eq.\ \eqref{sigma-I}. Note, however, that since the field's frequency, being in the vibrational regime, is much larger than the rotational frequency, we have $\frac{\omega^2 \bar\omega}{\bar\omega^2-\omega^2}\approx -\bar\omega$. As a result, this term $\sigma_I^{\rm rot}(\omega)$ becomes independent of $\omega$.

The quantity $S_{\nu_0}^{\nu_n}$ may be roughly estimated assuming that
the vibrational mode is purely harmonic, i.e., vibrational
wavefunctions are harmonic oscillator wavefunctions. The operator
$d^{(r)}_0$ may be written as $-\sum_{i=1}^Ls^{\parallel}_i+\zeta_NS$ where $s^{\parallel}_i$ is the
projection of $\mathbf{s}_i$ onto the molecular axis. The quantity
$\left(d^r_0\right)_{\nu_n}^{\nu_0}$ may be approximated by
$-f_{\nu_n}^{\nu_0}\bar{s}^{\parallel}+\zeta_N
S_{\nu_n}^{\nu_0}$ where $f_{\nu_n}^{\nu_0}=
\langle{\nu_n|\nu_0}\rangle$ is the Franck-Condon factor and
$\bar{s}^\parallel=\langle \mu_0\Sigma|\sum_{i=1}^L s^\parallel_i| \mu_0\Sigma \rangle$.
If one takes, approximately,
$f_{\nu_n}^{\nu_0}\approx\delta_{\nu_n}^{\nu_0}$ then the
electronic part drops out when summed over all $\nu_n \neq
\nu_0$. In this case, $\left(d^r_0\right)_{\nu_n}^{\nu_0}\approx
\zeta_NS_{\nu_n}^{\nu_0}$. Typically, $S_{\nu_n}^{\nu_0}$ decreases
rapidly as $\nu_n$ increases. Thus, only the term with
$\nu_n=\nu_0\pm 1$ 
makes dominant contribution to the sum in Eq.\ \eqref{Vibrational Sigma 2}.

\subsubsection{Screening of electronic-range-frequency fields}

In the case where $\omega$ is of the order of an electronic energy, $\omega\sim\omega_{\rm el}$, the dominant contributions to the scalar polarizability (\ref{alpha00}) come from 
the electronic terms (\ref{45a}) and (\ref{45b}). Due to the factor ${\cal M}_I$, the contribution $\beta_I$ to the suppression 
factor (\ref{sigmaI}) is also significant. Thus, in this case, we have
\begin{subequations}\label{Electronic Sigma 2}
\begin{eqnarray}
\sigma_I&=&\sigma_I^{\rm el}(\omega)+\sigma_I^{\rm vib}(\omega)+\sigma_I^{\rm rot}(\omega)\,,\label{sigma_el}\\
\sigma_I^{\rm el}(\omega)&\approx&  -\frac{\omega^2M_I}{3M_NZ_I}
 \left[\alpha^{\parallel}_{\rm el}(\omega)
 +2\alpha^{\perp}_{\rm el}(\omega)\right]\,,
\label{sigma_el+}
\end{eqnarray}
\end{subequations} 
where the quantities $\bar\omega$, $\bar S$, $\bar d$, $S_{\nu_0}^{\nu_n}$ and
$\left(d^r_0\right)_{\nu_n}^{\nu_0}$ are as given in Eqs.\ \eqref{56},\ \eqref{57},\ \eqref{58},\ \eqref{60} and\ \eqref{61}, respectively. We point out that the expression (\ref{sigma_el+}) is analogous to the screening factor for the oscillating electric field in atoms found in \cite{FlambaumOscillating}. 

The quantities $\sigma_I^{\rm rot}(\omega)$ and $\sigma_I^{\rm vib}(\omega)$ are defined in Eqs.\ \eqref{sigma-I} and\ \eqref{sigma_vib}, respectively. Note that just as in the vibrational energy regime, here, $\sigma_I^{\rm rot}(\omega)$ and $\sigma_I^{\rm vib}(\omega)$ is independent of $\omega$.

\subsubsection{Summary on different contributions to the screening coefficient}

In the rotational energy regime ($\omega \sim 10^{-5}$ a.u.), only $\sigma_I^{\rm rot}$ given by Eq.\ (\ref{sigma-I}) contributes to $\sigma_I$. As we move up to the vibrational energy regime ($\omega \sim 10^{-3}$ a.u.), $\sigma_I^{\rm vib}(\omega)$ defined in Eq.\ (\ref{60b}) becomes comparable to $\sigma_I^{\rm rot}$ and $\sigma_I^{\rm el}$, on the other hand, is still negligible. In the electronic energy regime ($\omega \sim 0.1$ a.u.), $\sigma_I^{\rm el}$ from Eq.\ (\ref{sigma_el+}) becomes comparable to its rotational and vibrational counterpart; now all three terms contribute to $\sigma_I$.

\begin{figure}[htb]
    \centering
    \includegraphics[scale=0.24]{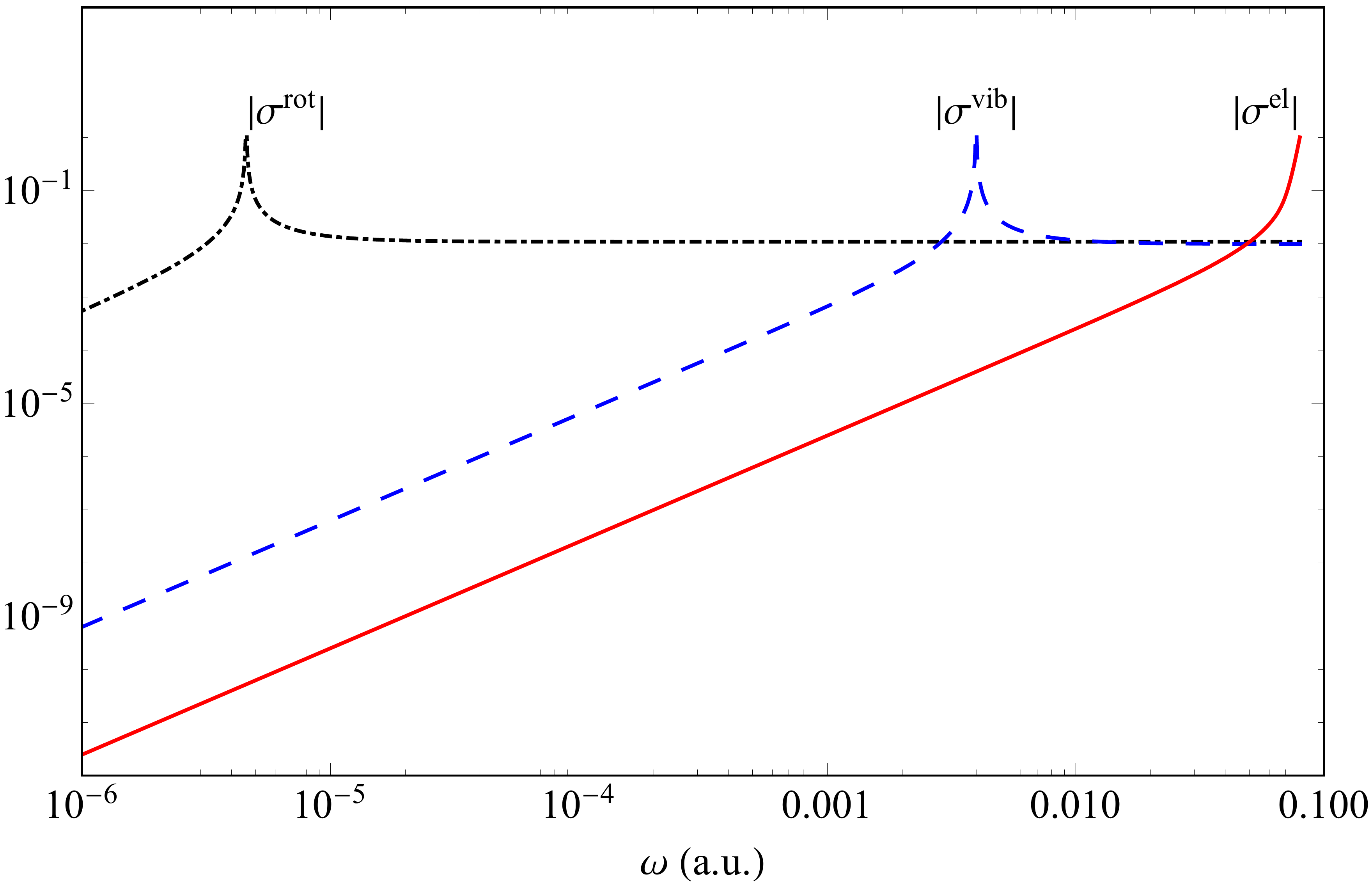}
    \caption{Comparison of the magnitudes of the three contributions to $\sigma_I$: $\sigma_I^{\rm rot}$, $\sigma_I^{\rm vib}$ and $\sigma_I^{\rm el}$ in  CaF molecule. The frequency $\omega$ is presented in atomic units. 
    For low frequency in the rotational regime, $\sigma_I^{\rm rot}$ dominates. For frequency in the vibrational regime, both $\sigma_I^{\rm rot}$ and $\sigma_I^{\rm vib}$ contribute; $\sigma_I^{\rm el}$ is still negligible. For large frequency in the electronic regime, each of the three terms is comparable to other two.}
    \label{Compare three terms}
\end{figure}

In Fig.\ \ref{Compare three terms}, we plot the behaviour of the three quantities $\sigma_I^{\rm rot}(\omega)$, $\sigma_I^{\rm vib}(\omega)$ and $\sigma_I^{\rm el}(\omega)$ as functions of frequency $\omega$. The change is the significance of these terms as one moves up the frequency scale is clearly demonstrated.

Note that each of the three plots shows a single resonance (the spikes in the case of $\sigma_I^{\rm rot}$ and $\sigma_I^{\rm vib}$, the upward tail of $\sigma_I^{\rm el}$) due to the applied approximations. 
In reality, $\sigma_I^{\rm rot}$ and $\sigma_I^{\rm vib}$ should have many more resonances. However, since we assumed that the molecule is in the ground state when the external field is absent and the contribution to $\sigma_I^{\rm vib}$ of all states with vibrational quantum number large than 1 is negligible, $\sigma_I^{\rm rot}$ and $\sigma_I^{\rm vib}$ each has, under these approximations, a single resonance. On the other hand, $\sigma_I^{\rm el}$ possesses many resonances which we do not include in Fig.\ \ref{Compare three terms} (these resonances are at frequencies higher than the range plotted in Fig.\ \ref{Compare three terms}). 

\subsection{Resonance enhancement from the lowest rotational transition}

In this subsection, we consider the resonance enhancement of an oscillating electric field in a diatomic molecule by the lowest rotational level. This case may be of interest for experimental applications because the energy of this level is in the microwave region and is thus quite accessible experimentally.

Since the $\overset{\scriptscriptstyle\leftrightarrow}{\boldsymbol{\gamma}_I}$ tensor contains only the scalar part $\gamma_I$ as in Eq.\ (\ref{48-}), the formula (\ref{36}) for the enhancement of the electric field may by rewritten as
\begin{equation}\label{SumUp2}
    E_I = \epsilon_I(\omega) E_0 \sin\omega t\,,
\end{equation}
where $\epsilon_I$ denotes the enhancement factor,
\begin{equation}
\label{64}
\epsilon_I(\omega)=\frac{\omega^2M_I \gamma_I}{\zeta_e Z_I M_T \Gamma}\,.
\end{equation}

When the frequency of the external field is equal to the lowest rotational state frequency (\ref{56}), $\omega=\bar\omega$, the equation (\ref{46}) yields 
$\gamma_I = -\frac23(-1)^I {\cal M}_I \bar S \bar d$, where $\bar S$ and $\bar d$ are given in Eqs.\ (\ref{57}) and (\ref{58}), respectively. The width of this transition may be approximated by the electric-dipole one-photon decay rate, $\Gamma \approx \frac43\bar{\omega}^3\bar{d}^2$. In this case, the enhancement factor (\ref{64}) reads (in the limit $m_e\ll M_I$)
\begin{equation}\label{Epsilon_I_BO}
\epsilon_I = -\left(-1\right)^I\frac{\mu_N\bar{S}}{2Z_I\bar{\omega}\bar{d}}\,.
\end{equation}
This factor gives a large enhancement of the electric field owing to the small quantity $\bar\omega$ in the denominator.

The formula\ \eqref{Epsilon_I_BO} is applicable only for sufficiently weak electric field, i.e., when the Rabi frequency $\Omega=\langle0|\mathbf{d}\cdot\mathbf{E}_0|n\rangle$ satisfies the condition $\Omega \ll \Gamma$. If this condition is not met, the enhancement of the electric field is described by the non-perturbative formula (\ref{With_Omega_Rabi}). In this case, instead of Eq.\ (\ref{Epsilon_I_BO}) we have a more general expression for the enhancement factor
\begin{equation}\label{Non_perturbative}
    \epsilon_I = -\left(-1\right)^I\frac{4\mu_N\bar{\omega}^5\bar S\bar{d}}{Z_I\left(8\bar{\omega}^6\bar{d}^2+9E_0^2\right)}\,.
\end{equation}
Note that Eq.\ \eqref{Non_perturbative} reduces to Eq.\ \eqref{Epsilon_I_BO} if $E_0\ll\bar{\omega}^3\bar d$. 

According to Eq.\ \eqref{E_crit}, the magnitude of the induced field, $E_I$, reaches its maximum at $E_0=E_{\rm crit}\equiv\sqrt{8}\bar{\omega}^3\bar{d}/3$. The value
$\epsilon_I(E_{\rm crit})$ is given by
\begin{equation}
     \epsilon_I(E_{\rm crit}) = -(-1)^I\frac{\mu_N\bar{S}}{4Z_I\bar{\omega}\bar{d}}\,,
\end{equation}
and the corresponding value of $E_I$ is  
\begin{equation}\label{E_I_max_Natural_width}
E_I^{\rm max} = -(-1)^I\frac{\mu_N\bar{\omega}^2\bar{S}}{3\sqrt{2}Z_I}\,.
\end{equation}

Note that here, for simplicity, we have assumed that $\Gamma_n$ is the natural width of the state $|n\rangle$. In reality, there may be other, possibly larger, contributions to $\Gamma_n$ such as Doppler width, collison width, etc. Nevertheless, as proved in Eq.\ \eqref{E_I_max}, the formula\ \eqref{E_I_max_Natural_width} for $E_I^{\rm max}$ has the same form regardless of which $\Gamma_n$ is assumed. The critical field $E_0$, however, depends linearly on $\Gamma_n$.

\subsection{Numerical results}

In this subsection, we present numerical estimates for the suppression and enhancement factors $\sigma_I$ and $\epsilon_I$ given by Eqs.\ (\ref{sigma-I}), (\ref{Vibrational Sigma 2}), (\ref{Electronic Sigma 2}), (\ref{Epsilon_I_BO}) and (\ref{Non_perturbative})
for various frequency regimes and different strengths of the external electric field. These estimates will be given for some simple diatomic molecules with well-studied polarizability properties: lithium hydride (LiH), sodium hydride (NaH), boron fluorine (BF) and calcium fluorine (CaF). 


For our numerical estimates, we take the values of the mean electric dipole parameter $\bar d$ from Refs.\ \cite{DipoleLiHNaH,DipoleBF,DipoleCaF}. The mean values of the inter-nuclear distance $\bar S$ 
for the molecules under consideration are available in the NIST database\ \citep{NIST}. The quantity $S_{\nu_0}^{\nu_n}$ may be roughly estimated assuming that the vibrational mode is purely harmonic. The quantity $\left(d^r_0\right)_{\nu_n}^{\nu_0}$ is approximately given by $\zeta_NS_{\nu_n}^{\nu_0}$. 

As noted above, the sums in Eqs.\ \eqref{Vibrational Sigma 2} and\ \eqref{Electronic Sigma 2} contain effectively one term corresponding to $\nu_0=0$ and $\nu_n=1$. The  values of the parallel $\alpha_{\rm el}^\parallel$ and perpendicular $\alpha_{\rm el}^\perp$ electronic polarizabilities for the molecules under inspection are presented in Ref.\ \cite{Polarizability}. 

For the estimates of the suppression factor $\sigma_I$ we consider three frequency regimes: low-frequency regime, where the frequency of the external field is taken to be half of the first rotational frequency, $\omega = \bar\omega/2$; intermediate-frequency regime with $\omega= \omega_e/2$, where $\omega_e$ is the ground state vibrational constant; electronic-transition range $\omega\approx \omega_{\rm el}$ where $\omega_{\rm el}$ is the lowest E1 electronic transition frequency. The value of $\omega_{e}$ may be found in the NIST database\ \cite{NIST}) while the electronic-range frequency $\omega_{\rm el}$ is available in Ref.\ \cite{Polarizability}. 

Recall that the index $I$ labels the nuclei in the diatomic molecule. Without loss of generality we assume that $I=1$ stands for the lighter nucleus while $I=2$ for the heavier one. The values of the suppression factors of the electric field at these nuclei in different frequency regimes are presented in Table\ \ref{TableResult_Sigma}. The values of the enhancement factors $\epsilon_1$ are given in Table\ \ref{TableResult_Epsilon} for a weak $E_0\ll E_{\rm crit}$ and strong $E_0= E_{\rm crit}$ external field in resonance with the first rotational level. Here $E_{\rm crit}=\sqrt{8}\bar{\omega}^3\bar{d}/3$ is a the value of the of the electric field $E_0$ at which the induced field $E_I$ reaches its maxium. Incidentally, $E_{\rm crit}$ also serves as a rough marker above which the general formula (\ref{Non_perturbative}) applies while under this value a simplified formula (\ref{Epsilon_I_BO}) gives sufficient accuracy. 
\begin{table}[ht]
{\begin{tabular}{@{}|c|c|c|c|c|c|c|c|c|@{}} \colrule
    \rule{0pt}{10pt} & $\bar{S}$ & $\bar{d}$ & $\begin{matrix}\bar{\omega} \\\left(\times 10^{-5}\right) \end{matrix}$ & $\begin{matrix}\omega_{e}\\\left(\times 10^{-3}\right)\end{matrix}$ & $\omega$ & $\sigma_1$ & $\sigma_2$ \\
    \colrule
    \rule{0pt}{10pt}\multirow{3}{*}{LiH} & \multirow{3}{*}{3.0} & \multirow{3}{*}{-2.3} & \multirow{3}{*}{6.9} & \multirow{3}{*}{6.4} & $3.4\times10^{-5}$ & -0.17 & 0.06 \\
   \cline{6-8}
   \rule{0pt}{10pt}& & & & & $3.2\times 10^{-3}$ & 0.70 & -0.23 \\
   \cline{6-8}
   \rule{0pt}{10pt}& & & & & 0.12 & 0.37 & 0.20 \\
    \colrule
    \rule{0pt}{10pt}\multirow{3}{*}{NaH} & \multirow{3}{*}{3.6} & \multirow{3}{*}{-2.6} & \multirow{3}{*}{4.5} & \multirow{3}{*}{5.3} & $2.2\times 10^{-5}$ & -0.16 & 0.02 \\
   \cline{6-8}
  \rule{0pt}{10pt} & & & & & $2.7\times 10^{-3}$ & 0.51 & -0.05 \\
   \cline{6-8}
   \rule{0pt}{10pt}& & & & & 0.11 & 0.35 & 0.23 \\
    \colrule
    \rule{0pt}{10pt}\multirow{3}{*}{BF}  &  \multirow{3}{*}{2.4} & \multirow{3}{*}{0.4} & \multirow{3}{*}{1.4} & \multirow{3}{*}{6.4} & $7.0\times 10^{-6}$ & 0.01 & -0.004 \\
    \cline{6-8}
   \rule{0pt}{10pt}& & & & & $3.2\times 10^{-3}$ & -0.02 & 0.01 \\
   \cline{6-8}
   \rule{0pt}{10pt}& & & & & 0.22 & 0.38 & 0.38 \\
    \colrule
    \rule{0pt}{10pt}\multirow{3}{*}{CaF} &  \multirow{3}{*}{3.7} & \multirow{3}{*}{1.2} & \multirow{3}{*}{0.5} & \multirow{3}{*}{2.6} & $2.5\times 10^{-6}$ & 0.01 & -0.004 \\
    \cline{6-8}
   \rule{0pt}{10pt}& & & & & $1.3\times 10^{-3}$ & -0.05 & 0.01 \\
   \cline{6-8}
   \rule{0pt}{10pt}& & & & & 0.075 & 1.4 & 1.3 \\
    \colrule
    \end{tabular}}
    \caption{Estimates of the suppression factors for the electric field on the lighter ($\sigma_1$) and heavier ($\sigma_2$) nuclei in the molecules LiH, NaH, BF and CaF.
    $\bar{S}$ is the ground state mean inter-nuclear distance, $\bar{d}$ is the
    ground state mean electric dipole, $\bar{\omega}$ is the rigid-rotor rotational
    energy with angular momentum $1$ and $\omega_e$ is the ground state first
    vibrational constant. All quantities are presented in atomic units\footnote{For the
    convenience of unit conversion, we recall that one atomic unit of length (Bohr) is equivalent
    to $0.529$ \AA, one atomic unit of electric dipole moment is equivalent to $0.393$
    Debye and one atomic unit of energy (Hartree) is equivalent to $27.21$ eV or
    $6.58\times 10^6$~GHz or $2.19\times 10^5{\ \rm cm}^{-1}$.
    The suppression factors $\sigma_{1,2}$ are, of course, dimensionless.}.}\label{TableResult_Sigma}
\end{table}

    \begin{table}[ht]
    \centering
    \begin{tabular}{@{}|c|c|c|c|c|@{}}
    \colrule
        & $\begin{matrix}E_{\rm crit}\\\left({\rm V/cm}\right)\end{matrix}$ & $\begin{matrix}\epsilon_1\\\left(E_0\ll E_{\rm crit}\right)\end{matrix}$ & $\begin{matrix}\epsilon_1\\\left(E_0=E_{\rm crit}\right)\end{matrix}$ & $\begin{matrix}E_1^{\rm max}\\\left({\rm V/cm}\right)\end{matrix}$ \\
        \colrule
        LiH      &      $3.7\times 10^{-3}$   & $-1.5\times10^{7}$ &     $-7.7\times10^{6}$       &  $-2.9\times10^{4}$ \\
        \colrule
        NaH      &      $1.1\times 10^{-3}$     & $-2.7\times10^{7}$ &    $-1.4\times10^{7}$      & $-1.5\times10^{4}$  \\
        \colrule
        BF       &      $5.3\times10^{-6}$         & $5.5\times10^{8}$ &     $2.8\times10^{8}$    & $1.5\times10^{3}$  \\
        \colrule
        CaF      &      $7.3\times10^{-7}$           & $1.1\times10^{9}$ &   $5.3\times10^{8}$     & $3.9\times10^{2}$ \\
        \colrule
        & $\begin{matrix}E_{\rm crit}\\\left({\rm V/cm}\right)\end{matrix}$ & $\begin{matrix}\epsilon_2\\\left(E_0\ll E_{\rm crit}\right)\end{matrix}$ & $\begin{matrix}\epsilon_2\\\left(E_0=E_{\rm crit}\right)\end{matrix}$ & $\begin{matrix}E_2^{\rm max}\\\left({\rm V/cm}\right)\end{matrix}$\\
        \colrule
        LiH      &      $3.7\times 10^{-3}$   &  $5.1\times10^{6}$  &   $2.6\times10^{6}$       &    $9.5\times10^{3}$     \\
        \colrule
        NaH      &      $1.1\times 10^{-3}$   &  $2.5\times10^{6}$  &   $1.2\times10^{6}$       &    $1.4\times10^{3}$      \\
        \colrule
        BF       &      $5.3\times10^{-6}$   &  $-3.1\times10^{8}$    &   $-1.5\times10^8$         &      $-8.2\times10^{2}$    \\
        \colrule
        CaF      &      $7.3\times10^{-7}$   &   $-2.7\times10^{8}$    &   $-1.3\times10^8$         &      $-9.6\times10^{1}$    \\      
        \colrule
    \end{tabular}
    \caption{Estimates of the enhancement factors $\epsilon_1$ and $\epsilon_2$ in the molecules LiH, NaH, BF and CaF.}
    \label{TableResult_Epsilon}
\end{table}

\section{Screening and enhancement of electric field in polyatomic molecules}\label{General Consideration}

In this section, we derive the analogs of the equations (\ref{Diatomic_Final_Result}) and (\ref{With_Omega_Rabi}) for the screening and resonant enhancement of alternating electric fields in polyatomic molecules. 

\subsection{The polyatomic molecule Hamiltonian in the center-of-mass frame }


In this subsection we consider the Hamiltonian of a polyatomic molecule in the center-of-mass frame. This subsection contains no new results and serves mainly to specify the notation for the next subsection. For a more extensive discussion of the separation of the nuclear and electronic motions in molecules, see, for example, Ref.\ \cite{Sutcliffe}.

Consider a molecule composed of $H$ (for ``heavy'') nuclei and $L$ (for ``light'') electrons in an external electric field $\mathbf{E}_{\rm ext}$. The masses and charges of the nuclei will be denoted by
$M_I$ and $Z_I$ respectively; the nuclear positions and momenta operators are $\mathbf{R}_I$ and  $\mathbf{P}_I$, respectively. The electrons positions and momenta are denoted by $\mathbf{r}_i$ and $\mathbf{p}_i$, respectively. Here we use capital letters $I,J=1,\ldots,H$ to label the nuclei, and
small letters are attributed to the electrons $i,j=1,\dots,L$.

The Hamiltonian of a polyatomic molecule has the standard form
\begin{equation}\label{General_Hamiltonian}
\begin{aligned}
  H_{\rm mol}&=K+{{V}_{0}}+V \,,\\
  K&=\sum\limits_{I=1}^{H}{\frac{\mathbf{P}_I^2}{2M_I}}+\sum\limits_{i=1}^{L}{\frac{\mathbf{p}_i^2}{2}} \,,\\
  V_0&=\sum\limits_{I<J}^{H}{\frac{{{Z}_{I}}{{Z}_{J}}}{{{R}_{IJ}}}} -\sum\limits_{I=1}^{H}{\sum\limits_{i=1}^{L}{\frac{{{Z}_{I}}}{{{R}_{Ii}}}}}+\sum\limits_{i<j}^{L}{\frac{1}{{{r}_{ij}}}}\,,\\
  V&=-{{\mathbf{E}}_{\rm ext}}\cdot \left(\sum\limits_{I=1}^{H}{{{Z}_{I}}{{\mathbf{R}}_{I}}} -\sum\limits_{i=1}^{L}{{{\mathbf{r}}_{i}}}\right) \,,
\end{aligned}
\end{equation}
where $R_{IJ}=|{\bf R}_I-{\bf R}_J|$, $R_{Ij}=|{\bf R}_I- {\bf r}_j|$ and $r_{ij}=|{\bf r}_i - {\bf r}_j|$. In this Hamiltonian, we ignore all spins of particles.

Following the same steps as in the case of diatomic molecules, Sect.\ \ref{SecIIA}, we introduce the parameters of total nuclear mass $M_N= \sum_{I=1}^H M_I$ and the total nuclear charges $Z_N= \sum_{I=1}^H Z_I$. The total molecular mass and charge are $M_T = M_N+L$, $Z_T = Z_N-L$, respectively. Using these notation, we perform a change of variables $({\bf R}_I, {\bf r}_i)\to ({\bf S}_T,{\bf S}_I,{\bf s}_i)$ similar to Eqs.\ (\ref{Diatomic_Coordinates_Transformation}):
\begin{subequations}
\label{General_Coordinates_Transformation}
\begin{eqnarray}
  {{\mathbf{S}}_{T}}&=&\sum\limits_{J=1}^{H}{U_{N}^{HJ}{{\mathbf{R}}_{J}}} + \frac{1}{{{M}_{T}}}\sum\limits_{i=1}^{L}{{{\mathbf{r}}_{i}}}\,,\label{69a}\\
   {{\mathbf{S}}_{I}}&=&\sum\limits_{J=1}^{H}{U_{N}^{IJ}{{\mathbf{R}}_{J}}}\qquad (I=1,\ldots H-1)  \,,\label{69b}\\
   {{\mathbf{s}}_{i}}&=&{{\mathbf{r}}_{i}}-\frac1{M_N}\sum\limits_{I=1}^{H}{{M}_{I}}{{\mathbf{R}}_{I}}
   \,,\label{69c}
\end{eqnarray}
\end{subequations}
where $U_N$ is an $H\times H$ invertible matrix whose bottom-row elements have the form $U_N^{HJ}={M_J}/{M_T}$. The inverse of $U_N$ is denoted by $U_N^{-1}$ and has elements on the last column of the form $\left(U_N^{-1}\right)^{IH}=M_T/M_N$.

The condition $U_N^{HJ}={M_J}/{M_T}$ means that the vector $\mathbf{S}$ in Eq.\ (\ref{69a}) is the coordinates of the molecule's center of mass. The submatrix $U_N^{IJ}$, with $I,J=1,\ldots, H-1$, in Eq.\ (\ref{69b}) specifies the relative coordinates of the nuclei in the center-of-mass frame. 
%
%
Physical quantities, such as the (expectation value of) total electric field at a nucleus in the molecule, must be independent of $U_N^{IJ}$. 

It should also be noted that the equation (\ref{69c}) defines the coordinates of electrons ${\bf s}_i$ with respect to the nuclear center of mass. 

The conjugated momenta for the coordinates $({\bf S}_T,{\bf S}_I,{\bf s}_i)$ are denoted by
$({\bf Q}_T,{\bf Q}_I,{\bf q}_i)$. The original momenta ${\bf P}_I$ and ${\bf p}_i$ may be expressed in terms of the new ones
%
\begin{subequations}\label{General_Momenta_Transformation}
\begin{eqnarray}
  {{\mathbf{p}}_{i}}&=&\mathbf{q}_i+\frac1{{M}_{T}}{{\mathbf{Q}}_{T}}\,,\\
  {{\mathbf{P}}_{I}}&=&\sum\limits_{J=1}^{H-1}{U_{N}^{JI}{{\mathbf{Q}}_{J}}} -\frac{{{M}_{I}}}{{{M}_{N}}}\sum\limits_{i=1}^{L}{{{\mathbf{q}}_{i}}}
   +\frac{{{M}_{I}}}{{{M}_{T}}}{{\mathbf{Q}}_{T}}\,.
   \label{General_Momenta_Transformation-b}
\end{eqnarray}
\end{subequations}

Upon this transformation the Hamiltonian\ \eqref{General_Hamiltonian} acquires the form\ \eqref{4} with the center-of-mass Hamiltonian $H_T$ given by Eq.\ \eqref{CM_Hamiltonian} and the relative motion Hamiltonian given by
%
\begin{subequations}\label{General_Transformed_Hamiltonian}
\begin{eqnarray}
H_{\rm rel}&=&H_0+V_{\rm rel}\,,\label{Hrel}\\
 H_0&=&\sum\limits_{i=1}^{L}{\frac{\mathbf{q}_{i}^{2}}{2{{\mu }_{e}}}}+\sum\limits_{i<j}^{L}{ \frac{{{\mathbf{q}}_{i}}{{\mathbf{q}}_{j}}}{{{M}_{N}}}}\nonumber\\
 &&+\frac12\sum\limits_{I,J=1}^{H-1}
 \left(\mu _{N}^{-1}\right)^{IJ}{{\mathbf{Q}}_{I}}{{\mathbf{Q}}_{J}} +V_0\,,
 \label{H0-general}
 \\
 V_{\rm rel}&=&-\mathbf{d}\cdot\mathbf{E}_{\rm ext}\,.\label{72c}
\end{eqnarray}
\end{subequations}

Here $\mu_e$ is the reduced electron mass\ \eqref{reduced-electron-mass} and $\mu^{-1}_N$ is the inverse reduced nuclear mass $H\times H$-matrix with elements
\begin{equation}
   \left(\mu^{-1}_N\right)^{IJ}=\sum\limits_{K=1}^{H}{M_{K}^{-1}U_{N}^{KI}U_{N}^{KJ}} \,.
\label{mu-general}
\end{equation}
It is straightforward to show that $\mu^{-1}_N$ has the following properties
\begin{equation}
\begin{aligned}
\left(\mu _{N}^{-1}\right)^{IH}&=\left(\mu _{N}^{-1}\right)^{HI}=0 \qquad (I=1,\ldots, H-1)\,,\\
\left(\mu _{N}^{-1}\right)^{HH}&=M_{T}^{-1}\,.
\end{aligned}
\end{equation}

The interaction potential (\ref{72c}) involves the molecule's electric dipole moment $\bf d$ with respect to the center of mass $\mathbf{S}_T$. This EDM is described by the analog of Eq.\ (\ref{def-d})
\begin{equation}
\label{mol-EDM}
\mathbf{d}=-\zeta_e\sum\limits_{i=1}^{L}{{{\mathbf{s}}_{i}}}+\sum\limits_{I=1}^{H-1}{\zeta_{N}^{I}{{\mathbf{S}}_{I}}}\,,
\end{equation}
where $\zeta_e$ is the reduced electron charge (\ref{def-zeta}) and $\zeta_N^I$ is the reduced nuclear charge
\begin{equation}
\label{zeta-general}
\zeta_{N}^{I}=\sum\limits_{J=1}^{H}{{{\left( U_{N}^{-1} \right)}^{JI}}{{Z}_{J}}} \qquad
(I=1,\ldots, H-1)\,.
\end{equation}





The representation\ \eqref{General_Transformed_Hamiltonian} allows us to apply perturbation theory with $V_{\rm rel}$ considered as a perturbation. The unperturbed wavefunctions $|n\rangle$ are defined with respect to the Hamiltonian $H_0$ given in Eq.\ \eqref{H0-general}, i.e., $H_0 |n\rangle = {\cal E}_n |n\rangle$. The evolution of the ground state $|0\rangle$ shall be described by the wavefunction\ \eqref{20} (in the off resonance case) or\ \eqref{psi+Gamma}. We will use these wavefunctions to find the expectation value of the operator of electric field at the $I^{\rm th}$ nucleus.

\subsection{Off-resonance screening of an oscillating external electric field}

Let us assume that the frequency $\omega$ of the oscillating electric field\ \eqref{Eext} is far from any molecular transition frequency $\omega_{n0}$. When the magnitude of this field is sufficiently weak, the time-dependent perturbation theory may be applied to compute the expectation value of the operator of electric field at the $I^{\rm th}$ nucleus (\ref{Diatomic_E'_2}). This procedure is the same as presented in Sect.\ \ref{SectIIc}. As a result, for the electric field at $I^{\rm th}$ nucleus we arrive at the same expression\ \eqref{E_I}
\begin{equation}\label{General_E_I+}
\begin{aligned}
\langle {\bf E}_I\rangle & = \frac{M_I Z_T}{M_T Z_I} {\bf E}_0\cos\omega t\\
&+\frac{2\omega^2 \cos\omega t}{Z_I}
 \sum_n \frac{{\rm Im}\, \left[ \langle 0 |\boldsymbol{\Pi}_I | n\rangle \langle n |{\bf d}\cdot {\bf
E}_0|0\rangle\right]}{\omega_{n0}^2 - \omega^2}\,,
\end{aligned}
\end{equation}
where $\boldsymbol{\Pi}_I$ is the truncated momentum operator of $I^{\rm th}$ nucleus. This operator is defined by the expression (\ref{General_Momenta_Transformation-b}) with the center-of-mass term removed, namely
\begin{equation}\label{General_Pi_I}
    \boldsymbol{\Pi}_I\equiv {\bf P}_I - \frac{{{M}_{I}}}{{{M}_{T}}}{{\mathbf{Q}}_{T}}
    =\sum\limits_{J=1}^{H-1}{U_{N}^{JI}{{\mathbf{Q}}_{J}}} -\frac{{{M}_{I}}}{{{M}_{N}}}\sum\limits_{i=1}^{L}{{{\mathbf{q}}_{i}}}\,.
\end{equation}

Using the expressions of the Hamiltonian $H_0$ in Eq.\ (\ref{H0-general}) and molecular EDM operator (\ref{mol-EDM}), it is possible to show that the operator (\ref{General_Pi_I}) may be represented in the form
\begin{equation}
\begin{aligned}\label{General_Pi_I_Commutator}
  \boldsymbol{\Pi}_I
  =\frac{i{{M}_{I}}}{{{\zeta }_{e}}{{M}_{T}}}\left[-\mathbf{d}+\sum\limits_{J=1}^{H-1}{{{\mathcal{M} }^{IJ}}{{\mathbf{S}}_{J}}},H_0 \right]\,,
\end{aligned}
\end{equation}
where
\begin{equation}
\label{M-IJ}
    {{\mathcal{M} }^{IJ}}=\zeta _{N}^{J}-\left( {{M}_{N}}+{{Z}_{N}} \right){{\left( U_{N}^{-1} \right)}^{IJ}}
\end{equation}
is a generalization of the quantity $\mathcal{M}_I$ in Eq.\ (\ref{calM}).

Substituting Eq.\ \eqref{General_Pi_I_Commutator} into Eq.\ \eqref{General_E_I+}, one finds that the electric field at $I^{\rm th}$ nucleus is given by the expression similar to Eq.\ (\ref{Diatomic_Final_Result}),
\begin{equation}\label{General_Final_Result}
\left\langle {{\mathbf{E}}_{I}} \right\rangle = \left[ \frac{{{M}_{I}}{{Z}_{T}}}{{{M}_{T}}{{Z}_{I}}}-\frac{{{\omega }^{2}}{{M}_{I}}}{{\zeta }_{e}{M_T{Z}_{I}}}\left(\overset{\scriptscriptstyle\leftrightarrow}{\boldsymbol{\alpha }} -\overset{\scriptscriptstyle\leftrightarrow}{\boldsymbol{\beta }}_I\right)\right]{{\mathbf{E}}_{0}}\cos \omega t\,,
\end{equation}
where $\overset{\scriptscriptstyle\leftrightarrow}{\boldsymbol{\alpha }}$ is the molecule's polarizability tensor 
\begin{equation}\label{polyatomic_Polarizability}
\overset{\scriptscriptstyle\leftrightarrow}{\boldsymbol{\alpha
}}=2\sum\limits_{n} \frac{\omega_{n0}}{\omega
_{n0}^{2}-\omega^2}\langle 0|\mathbf{d}|n\rangle
\langle n|\mathbf{d}|0\rangle\,,
\end{equation}
and $\overset{\scriptscriptstyle\leftrightarrow}{\boldsymbol{\beta
}}_I$ is the polyatomic analog of the tensor (\ref{Diatomic_Beta}),
\begin{equation}\label{General_Beta}
\overset{\scriptscriptstyle\leftrightarrow}{\boldsymbol{\beta
}}_I=2\sum\limits_{n}\frac{{{\omega }_{n0}}}{\omega
_{n0}^{2}-{{\omega}^{2}}} \langle 0|\sum\limits_{J=1}^{H-1}\mathcal{M}^{IJ}\mathbf{S}_J|n\rangle\langle n|\mathbf{d}|0\rangle\,.
\end{equation}

We point out that the results (\ref{General_Final_Result})--(\ref{General_Beta}) reduce to the analogous results (\ref{Diatomic_Final_Result})--(\ref{Diatomic_Beta}) for diatomic molecules given in Sect.\ \ref{SectIIc} upon the special choice of the matrix 
$U_N=M_T^{-1}\left( 
\begin{array}{cc}
M_T & -M_T \\ M_1 & M_2
\end{array}
\right)$.


It is important to note that the value of the electric field at nucleus (\ref{General_Final_Result}) is independent of the choice of the matrix $U_N$ in Eq.\ (\ref{General_Coordinates_Transformation}). To demonstrate this, it is sufficient to prove that the tensors (\ref{polyatomic_Polarizability}) and (\ref{General_Beta}) are independent of this matrix.

As follows from Eqs.\ (\ref{General_Momenta_Transformation-b}) and (\ref{mu-general}), the Hamiltonian (\ref{H0-general}) is independent of the matrix $U_N$ when expressed in terms of the coordinates $\mathbf{r}_i$ and $\mathbf{R}_I$ and 
momenta ${\bf p}_i$ and ${\bf P}_I$. Thus, the eigenstates $|n\rangle$ of this Hamiltonian are also independent of the matrix $U_N$ when expressed in terms of the coordinates $\mathbf{r}_i$ and $\mathbf{R}_I$. Taking into account Eqs.\ (\ref{69b}) and (\ref{zeta-general}) one can easily see that the molecular EDM (\ref{mol-EDM}) is also independent of the matrix $U_N$ in the coordinates $\mathbf{r}_i$ and $\mathbf{R}_I$. Since these coordinates are integrated out in the matrix elements in Eq.\ (\ref{polyatomic_Polarizability}), we conclude that the molecular polarizability tensor $\overset{\scriptscriptstyle\leftrightarrow}{\boldsymbol{\alpha }}$ is independent of the choice of the matrix $U_N$.


Using Eqs.\ (\ref{69b}), (\ref{zeta-general}) and (\ref{M-IJ}) one may check the edientity $\sum\limits_{J=1}^{H-1}{{{\mathcal{M} }^{IJ}}{{\mathbf{S}}_{J}}}=\sum\limits_{J=1}^{H}{\left( {{M}_{J}}+{{Z}_{J}} \right){{\mathbf{R}}_{J}}}-\left( {{M}_{N}}+{{Z}_{N}} \right){{\mathbf{R}}_{I}}$. Thus, the quantity $\sum\limits_{J=1}^{H-1}{{{\mathcal{M} }^{IJ}}{{\mathbf{S}}_{J}}}$ is independent of the matrix $U_N$ in the coordinates $\mathbf{r}_i$ and $\mathbf{R}_I$. As a corollary, all matrix elements in Eq.\ (\ref{General_Beta}) are independent of the matrix $U_N$. This completes the proof that the expression for the electric field at nucleus (\ref{General_Final_Result}) is independent of the choice of the matrix $U_N$. 

The independence of the result\ \eqref{General_Final_Result} on $U_N$ means that one is free to define the relative nuclear coordinates in any convenient way.
Once a particular matrix $U_N$ is chosen, one can then solve for the states $n$ and proceed to calculating the fields as in Eq.\ \eqref{General_Final_Result}. The results one obtains this way will be the same as those obtained if some different $U_N$ was chosen.

\subsection{Resonance enhancement of an oscillating electric field}

Let us now assume that the frequency of the external electric field is in resonance with one of the molecular transition frequency, $\omega=\omega_{n0}$, associated with some excited state $|n\rangle$ with width $\Gamma_n$. Theoretically, due to the resonance, the magnitude of the electric field at nucleus infinitely grows with time if one discards the spontaneous decay rate of the excited state. Physically, this field can grow only up to the lifetime of the excited state, $\tau=1/\Gamma_n$. Therefore, the resonance enhancement of the electric field in the molecule is due to the factor $1/\Gamma_n$ which is large for lowest rotational and vibrational states. The off-resonance states, however, provide partial screening of the electric field in the same way as is described in Sect.\ \ref{Sect2D}. We stress that Eq.\ (\ref{35}) describing the resonance enhancement of the electric field in diatomic molecules holds for polyatomic molecules as well. Indeed, the derivation of this equation is purely formal and is not limited to the diatomic case.

Since the lowest ro-vibrational states in molecules possess very large lifetime, the resonance enhancement due to the factor $1/\Gamma_n$ becomes significantly larger than the screening due to the off-resonance states. In this case, the electric field is described by the analog of Eq.\ (\ref{36})
\begin{equation}
\label{36+}
\langle {\bf E}_I \rangle \approx
\frac{\omega^2m_eM_I}{\zeta_ee^2 Z_I M_T \Gamma_n}
\overset{\scriptscriptstyle\leftrightarrow}{\boldsymbol{\gamma }}_I
{\bf E}_0 \sin\omega t\,,
\end{equation}
where the tensor $\overset{\scriptscriptstyle\leftrightarrow}{\boldsymbol{\gamma
}}_I$ reads now
\begin{equation}
\label{General_gamma-tensor}
\overset{\scriptscriptstyle\leftrightarrow}{\boldsymbol{\gamma
}}_I = 2\langle 0| \mathbf{d} |n\rangle \langle n| \mathbf{d} |0\rangle 
-2\langle 0|\sum\limits_{J=1}^{H-1}\mathcal{M}^{IJ}\mathbf{S}_J|n\rangle \langle
n |{\bf d}|0\rangle\,.
\end{equation}
Here $\mathcal{M}^{IJ}$ is given in Eq.\ (\ref{M-IJ}).

We point out that the expression (\ref{36+}) for the enhancement of the electric field at nucleus is valid for a weak external field since it is obtained in perturbation theory. The condition of the applicability of perturbation theory reads $\Omega \ll \Gamma_n$, where $\Omega = |\langle 0|\mathbf{d}\cdot \mathbf{E}_0|n\rangle|$ is the Rabi frequency. For a stronger electric field one is to apply a more general formula derived non-perturbatively in \cite{SamsonovResonant}:
\begin{equation}
\langle {\bf E}_I \rangle =
\frac{\omega^2m_eM_I}{\zeta_ee^2 Z_I M_T }\frac{\Gamma_n}{\Gamma_n^2+2\Omega^2}
\overset{\scriptscriptstyle\leftrightarrow}{\boldsymbol{\gamma }}_I
{\bf E}_0 \sin\omega t\,.
\end{equation}

This formula has the same form as in the case of diatomic molecules (\ref{With_Omega_Rabi}) except for the tensor $\overset{\scriptscriptstyle\leftrightarrow}{\boldsymbol{\gamma
}}_I$ given by Eq.\ (\ref{General_gamma-tensor}). As a result, the formulae for the critical field $E_{\rm crit}$ and the maximal field $E^{\rm max}_I$ are the same as given in Eqs.\ \eqref{E_crit} and\ \eqref{E_I_max} but with $\overset{\scriptscriptstyle\leftrightarrow}{\boldsymbol{\gamma
}}_I$ given by Eq.\ (\ref{General_gamma-tensor}).


\section{Summary and discussion}\label{Conclusion}

In this paper, we derived the general formulae for the screening (\ref{full}) and resonance enhancement (\ref{With_Omega_Rabi}) of an oscillating electric field at a nucleus in a diatomic molecule. 

The screening formula\ \eqref{full} applies when the frequency of external electric field is far from any transition frequency in the molecule. This formula may be considered, on the one hand, as a generalization of the screening of a static electric field in molecules \cite{FlambaumMolecule} and, on the other hand, as an extension of the screening formula of oscillating electric field in atoms \cite{FlambaumOscillating}. For molecules, the screening of electric field exhibits some important features. Similarly to the atomic case \cite{FlambaumOscillating}, Eq.\ (\ref{full}) contains the term with dynamical polarizability $\overset{\scriptscriptstyle\leftrightarrow}{\boldsymbol{\alpha
}}$. However, there are extra terms (described by the tensor $\overset{\scriptscriptstyle\leftrightarrow}{\boldsymbol{\beta}}$) in Eq.\ (\ref{full}), which have no analogs in the atomic case. 

To uncover the role of the new terms in Eq.\ (\ref{full}), we considered the examples of some diatomic molecules in the electric field in different frequency regimes. We found that when the frequency of the electric field is of order of the frequency of ro-vibrational transitions in a molecule, the $\overset{\scriptscriptstyle\leftrightarrow}{\boldsymbol{\beta}}$-tensor in Eq.\ (\ref{full}) gives dominant contribution to the resulting electric field at nucleus since it is enhanced by the ratio $M_I/m_e$. This parameter makes the screening of the low-frequency electric field in molecules very different from that in atoms.

When the frequency of the external field approaches the energies of electronic transitions, the contributions from both tensors $\overset{\scriptscriptstyle\leftrightarrow}{\boldsymbol{\alpha}}$ and $\overset{\scriptscriptstyle\leftrightarrow}{\boldsymbol{\beta}}$ become significant. Thus, the molecules exhibit different screening behaviour in different frequency ranges. The typical dependence on the frequency of different contributions to the suppression coefficient is presented at Fig.\ \ref{Compare three terms}. In particular, the screening of the field in molecules in the microwave regime appears not as strong as it is in atoms \cite{FlambaumOscillating}. The summary of suppression coefficients $\sigma_I$ for some molecules in various frequency regimes is given in Table \ref{TableResult_Sigma}.

When the frequency of the external electric field approaches one of the transition frequency, the resonant excited state in the molecule is responsible for the linear-in-time growth of the electric field at nucleus up to the life-time of the excited state. The off-resonant states, however, provide partial screening of this electric field as is shown in Eq.\ (\ref{35}). The resonance enhancement may be up to the factor $10^9$ due to a small width of the state $\Gamma$, so it becomes dominating over the suppression factors as in Eq.\ (\ref{36}). 

The advantage molecules have over atoms is that they possess ro-vibrational states with energies in the microwave or even radio-frequency region. These states in molecules typically have very narrow spectral lines that makes the resonance enhancement of the oscillating electric field at nucleus very large. 

However, we caution the readers against using the resonance enhancement formula (\ref{36}) for the electric field at nucleus: Since this result is derived within the perturbatie approach, it is applicable only for sufficiently weak electric field under the constraint (\ref{contraint-omega}). For a stronger electric field, one should apply the non-perturbative formula (\ref{With_Omega_Rabi}) which was derived in \cite{SamsonovResonant}. Physically, this formula tells us that even in the resonance one cannot produce the electric field at nucleus stronger than the Coulomb field inside the molecule. The point of the perturbative formula (\ref{35}) is that one can take a very weak electric field in resonance with the molecular transition to produce sufficiently large oscillating electric field at nucleus. Some particular examples of such weak fields and their amplification factors $\epsilon_I$ are presented in Table \ref{TableResult_Epsilon}.

Although our main results (\ref{full}) and (\ref{With_Omega_Rabi}) are derived for the case of diatomic molecules, we present a generalization of these formulae to the polyatomic molecules in Sect. \ref{General Consideration}.

The results of this paper may have various physical applications since they represent a way out from the shielding of a static electric field at nucleus due to the Schiff theorem \cite{Schiff}. In particular, it would be interesting to develop a technique for measuring nuclear EDM using an oscillating electric field in resonance with molecular transition. Many diatomic molecules possess $\Omega$-doubling of states with splitting of order of 100 MHz. Such molecules have already proved useful for measuring electron's EDM, see, e.g., \cite{Baron:2013eja,Eckel2013,Bickman2009}. One can apply the electric field in resonance with this transition to induce a large electric field at nucleus which may be used to measure nuclear EDM. Another use of the results of this paper may be related to application of laser beams to stimulate nuclear transitions such as the neutron capture in the $^{139}$La nucleus proposed in \cite{Zaretskii79,Zaretskii81,Dzyublik92}. These issues deserve separate studies.



\section*{Acknowledgments}
This work is supported by the Australian Research Council Grant No. DP150101405 and by a Gutenberg Fellowship.

\bibliography{SchiffBib}

\begin{thebibliography}{29}%
\makeatletter
\providecommand \@ifxundefined [1]{%
 \@ifx{#1\undefined}
}%
\providecommand \@ifnum [1]{%
 \ifnum #1\expandafter \@firstoftwo
 \else \expandafter \@secondoftwo
 \fi
}%
\providecommand \@ifx [1]{%
 \ifx #1\expandafter \@firstoftwo
 \else \expandafter \@secondoftwo
 \fi
}%
\providecommand \natexlab [1]{#1}%
\providecommand \enquote  [1]{``#1''}%
\providecommand \bibnamefont  [1]{#1}%
\providecommand \bibfnamefont [1]{#1}%
\providecommand \citenamefont [1]{#1}%
\providecommand \href@noop [0]{\@secondoftwo}%
\providecommand \href [0]{\begingroup \@sanitize@url \@href}%
\providecommand \@href[1]{\@@startlink{#1}\@@href}%
\providecommand \@@href[1]{\endgroup#1\@@endlink}%
\providecommand \@sanitize@url [0]{\catcode `\\12\catcode `\$12\catcode
  `\&12\catcode `\#12\catcode `\^12\catcode `\_12\catcode `\%12\relax}%
\providecommand \@@startlink[1]{}%
\providecommand \@@endlink[0]{}%
\providecommand \url  [0]{\begingroup\@sanitize@url \@url }%
\providecommand \@url [1]{\endgroup\@href {#1}{\urlprefix }}%
\providecommand \urlprefix  [0]{URL }%
\providecommand \Eprint [0]{\href }%
\providecommand \doibase [0]{http://dx.doi.org/}%
\providecommand \selectlanguage [0]{\@gobble}%
\providecommand \bibinfo  [0]{\@secondoftwo}%
\providecommand \bibfield  [0]{\@secondoftwo}%
\providecommand \translation [1]{[#1]}%
\providecommand \BibitemOpen [0]{}%
\providecommand \bibitemStop [0]{}%
\providecommand \bibitemNoStop [0]{.\EOS\space}%
\providecommand \EOS [0]{\spacefactor3000\relax}%
\providecommand \BibitemShut  [1]{\csname bibitem#1\endcsname}%
\let\auto@bib@innerbib\@empty
\bibitem [{\citenamefont {Chupp}\ \emph {et~al.}(2017)\citenamefont {Chupp},
  \citenamefont {Fierlinger}, \citenamefont {Ramsey-Musolf},\ and\
  \citenamefont {Singh}}]{Chupp}%
  \BibitemOpen
  \bibfield  {author} {\bibinfo {author} {\bibfnamefont {T.}~\bibnamefont
  {Chupp}}, \bibinfo {author} {\bibfnamefont {P.}~\bibnamefont {Fierlinger}},
  \bibinfo {author} {\bibfnamefont {M.}~\bibnamefont {Ramsey-Musolf}}, \ and\
  \bibinfo {author} {\bibfnamefont {J.}~\bibnamefont {Singh}},\ }\href@noop {}
  {\  (\bibinfo {year} {2017})},\ \Eprint {http://arxiv.org/abs/1710.02504}
  {arXiv:1710.02504 [physics.atom-ph]} \BibitemShut {NoStop}%
\bibitem [{\citenamefont {Yamanaka}\ \emph {et~al.}(2017)\citenamefont
  {Yamanaka}, \citenamefont {Sahoo}, \citenamefont {Yoshinaga}, \citenamefont
  {Sato}, \citenamefont {Asahi},\ and\ \citenamefont {Das}}]{Yamanaka}%
  \BibitemOpen
  \bibfield  {author} {\bibinfo {author} {\bibfnamefont {N.}~\bibnamefont
  {Yamanaka}}, \bibinfo {author} {\bibfnamefont {B.~K.}\ \bibnamefont {Sahoo}},
  \bibinfo {author} {\bibfnamefont {N.}~\bibnamefont {Yoshinaga}}, \bibinfo
  {author} {\bibfnamefont {T.}~\bibnamefont {Sato}}, \bibinfo {author}
  {\bibfnamefont {K.}~\bibnamefont {Asahi}}, \ and\ \bibinfo {author}
  {\bibfnamefont {B.~P.}\ \bibnamefont {Das}},\ }\href {\doibase
  10.1140/epja/i2017-12237-2} {\bibfield  {journal} {\bibinfo  {journal} {Eur.
  Phys. J.}\ }\textbf {\bibinfo {volume} {A53}},\ \bibinfo {pages} {54}
  (\bibinfo {year} {2017})},\ \Eprint {http://arxiv.org/abs/1703.01570}
  {arXiv:1703.01570 [hep-ph]} \BibitemShut {NoStop}%
\bibitem [{\citenamefont {Safronova}\ \emph {et~al.}(2018)\citenamefont
  {Safronova}, \citenamefont {Budker}, \citenamefont {DeMille}, \citenamefont
  {Kimball}, \citenamefont {Derevianko},\ and\ \citenamefont
  {Clark}}]{Safronova}%
  \BibitemOpen
  \bibfield  {author} {\bibinfo {author} {\bibfnamefont {M.~S.}\ \bibnamefont
  {Safronova}}, \bibinfo {author} {\bibfnamefont {D.}~\bibnamefont {Budker}},
  \bibinfo {author} {\bibfnamefont {D.}~\bibnamefont {DeMille}}, \bibinfo
  {author} {\bibfnamefont {D.~F.~J.}\ \bibnamefont {Kimball}}, \bibinfo
  {author} {\bibfnamefont {A.}~\bibnamefont {Derevianko}}, \ and\ \bibinfo
  {author} {\bibfnamefont {C.~W.}\ \bibnamefont {Clark}},\ }\href {\doibase
  10.1103/RevModPhys.90.025008} {\bibfield  {journal} {\bibinfo  {journal}
  {Rev. Mod. Phys.}\ }\textbf {\bibinfo {volume} {90}},\ \bibinfo {pages}
  {025008} (\bibinfo {year} {2018})},\ \Eprint
  {http://arxiv.org/abs/1710.01833} {arXiv:1710.01833 [physics.atom-ph]}
  \BibitemShut {NoStop}%
\bibitem [{\citenamefont {Schiff}(1963)}]{Schiff}%
  \BibitemOpen
  \bibfield  {author} {\bibinfo {author} {\bibfnamefont {L.~I.}\ \bibnamefont
  {Schiff}},\ }\href {\doibase 10.1103/PhysRev.132.2194} {\bibfield  {journal}
  {\bibinfo  {journal} {Phys. Rev.}\ }\textbf {\bibinfo {volume} {132}},\
  \bibinfo {pages} {2194} (\bibinfo {year} {1963})}\BibitemShut {NoStop}%
\bibitem [{\citenamefont {Sandars}(1967)}]{Sandars1}%
  \BibitemOpen
  \bibfield  {author} {\bibinfo {author} {\bibfnamefont {P.~G.~H.}\
  \bibnamefont {Sandars}},\ }\href {\doibase 10.1103/PhysRevLett.19.1396}
  {\bibfield  {journal} {\bibinfo  {journal} {Phys. Rev. Lett.}\ }\textbf
  {\bibinfo {volume} {19}},\ \bibinfo {pages} {1396} (\bibinfo {year}
  {1967})}\BibitemShut {NoStop}%
\bibitem [{\citenamefont {Hinds}\ and\ \citenamefont
  {Sandars}(1980)}]{Sandars2}%
  \BibitemOpen
  \bibfield  {author} {\bibinfo {author} {\bibfnamefont {E.~A.}\ \bibnamefont
  {Hinds}}\ and\ \bibinfo {author} {\bibfnamefont {P.~G.~H.}\ \bibnamefont
  {Sandars}},\ }\href {\doibase 10.1103/PhysRevA.21.471} {\bibfield  {journal}
  {\bibinfo  {journal} {Phys. Rev. A}\ }\textbf {\bibinfo {volume} {21}},\
  \bibinfo {pages} {471} (\bibinfo {year} {1980})}\BibitemShut {NoStop}%
\bibitem [{\citenamefont {Sushkov}\ \emph {et~al.}(1984)\citenamefont
  {Sushkov}, \citenamefont {Flambaum},\ and\ \citenamefont
  {Khriplovich}}]{SFK1984}%
  \BibitemOpen
  \bibfield  {author} {\bibinfo {author} {\bibfnamefont {O.~P.}\ \bibnamefont
  {Sushkov}}, \bibinfo {author} {\bibfnamefont {V.~V.}\ \bibnamefont
  {Flambaum}}, \ and\ \bibinfo {author} {\bibfnamefont {I.~B.}\ \bibnamefont
  {Khriplovich}},\ }\href@noop {} {\bibfield  {journal} {\bibinfo  {journal}
  {Zh. Exp. Teor. Fiz.}\ }\textbf {\bibinfo {volume} {87}},\ \bibinfo {pages}
  {1521} (\bibinfo {year} {1984})}\BibitemShut {NoStop}%
\bibitem [{\citenamefont {Flambaum}\ \emph {et~al.}(1985)\citenamefont
  {Flambaum}, \citenamefont {Khriplovich},\ and\ \citenamefont
  {Sushkov}}]{FLAMBAUM1985}%
  \BibitemOpen
  \bibfield  {author} {\bibinfo {author} {\bibfnamefont {V.}~\bibnamefont
  {Flambaum}}, \bibinfo {author} {\bibfnamefont {I.}~\bibnamefont
  {Khriplovich}}, \ and\ \bibinfo {author} {\bibfnamefont {O.}~\bibnamefont
  {Sushkov}},\ }\href {\doibase https://doi.org/10.1016/0370-2693(85)90908-6}
  {\bibfield  {journal} {\bibinfo  {journal} {Phys. Lett. B}\ }\textbf
  {\bibinfo {volume} {162}},\ \bibinfo {pages} {213 } (\bibinfo {year}
  {1985})}\BibitemShut {NoStop}%
\bibitem [{\citenamefont {Flambaum}\ \emph {et~al.}(1986)\citenamefont
  {Flambaum}, \citenamefont {Khriplovich},\ and\ \citenamefont
  {Sushkov}}]{FLAMBAUM1986}%
  \BibitemOpen
  \bibfield  {author} {\bibinfo {author} {\bibfnamefont {V.}~\bibnamefont
  {Flambaum}}, \bibinfo {author} {\bibfnamefont {I.}~\bibnamefont
  {Khriplovich}}, \ and\ \bibinfo {author} {\bibfnamefont {O.}~\bibnamefont
  {Sushkov}},\ }\href {\doibase https://doi.org/10.1016/0375-9474(86)90331-3}
  {\bibfield  {journal} {\bibinfo  {journal} {Nuclear Physics A}\ }\textbf
  {\bibinfo {volume} {449}},\ \bibinfo {pages} {750 } (\bibinfo {year}
  {1986})}\BibitemShut {NoStop}%
\bibitem [{\citenamefont {Dzuba}\ \emph {et~al.}(1986)\citenamefont {Dzuba},
  \citenamefont {Flambaum}, \citenamefont {Silvestrov},\ and\ \citenamefont
  {Sushkov}}]{FlambaumConstant}%
  \BibitemOpen
  \bibfield  {author} {\bibinfo {author} {\bibfnamefont {V.}~\bibnamefont
  {Dzuba}}, \bibinfo {author} {\bibfnamefont {V.}~\bibnamefont {Flambaum}},
  \bibinfo {author} {\bibfnamefont {P.}~\bibnamefont {Silvestrov}}, \ and\
  \bibinfo {author} {\bibfnamefont {O.}~\bibnamefont {Sushkov}},\ }\href
  {\doibase https://doi.org/10.1016/0375-9601(86)90251-3} {\bibfield  {journal}
  {\bibinfo  {journal} {Phys. Lett. A}\ }\textbf {\bibinfo {volume} {118}},\
  \bibinfo {pages} {177 } (\bibinfo {year} {1986})}\BibitemShut {NoStop}%
\bibitem [{\citenamefont {Flambaum}(2018)}]{FlambaumOscillating}%
  \BibitemOpen
  \bibfield  {author} {\bibinfo {author} {\bibfnamefont {V.~V.}\ \bibnamefont
  {Flambaum}},\ }\href {\doibase 10.1103/PhysRevA.98.043408} {\bibfield
  {journal} {\bibinfo  {journal} {Phys. Rev. A}\ }\textbf {\bibinfo {volume}
  {98}},\ \bibinfo {pages} {043408} (\bibinfo {year} {2018})}\BibitemShut
  {NoStop}%
\bibitem [{\citenamefont {Dzuba}\ \emph {et~al.}(2018)\citenamefont {Dzuba},
  \citenamefont {Berengut}, \citenamefont {Ginges},\ and\ \citenamefont
  {Flambaum}}]{DBGF}%
  \BibitemOpen
  \bibfield  {author} {\bibinfo {author} {\bibfnamefont {V.~A.}\ \bibnamefont
  {Dzuba}}, \bibinfo {author} {\bibfnamefont {J.~C.}\ \bibnamefont {Berengut}},
  \bibinfo {author} {\bibfnamefont {J.~S.~M.}\ \bibnamefont {Ginges}}, \ and\
  \bibinfo {author} {\bibfnamefont {V.~V.}\ \bibnamefont {Flambaum}},\ }\href
  {\doibase 10.1103/PhysRevA.98.043411} {\bibfield  {journal} {\bibinfo
  {journal} {Phys. Rev. A}\ }\textbf {\bibinfo {volume} {98}},\ \bibinfo
  {pages} {043411} (\bibinfo {year} {2018})}\BibitemShut {NoStop}%
\bibitem [{\citenamefont {Flambaum}\ and\ \citenamefont
  {Samsonov}(2018)}]{SamsonovResonant}%
  \BibitemOpen
  \bibfield  {author} {\bibinfo {author} {\bibfnamefont {V.~V.}\ \bibnamefont
  {Flambaum}}\ and\ \bibinfo {author} {\bibfnamefont {I.~B.}\ \bibnamefont
  {Samsonov}},\ }\href {\doibase 10.1103/PhysRevA.98.053437} {\bibfield
  {journal} {\bibinfo  {journal} {Phys. Rev.}\ }\textbf {\bibinfo {volume}
  {A98}},\ \bibinfo {pages} {053437} (\bibinfo {year} {2018})},\ \Eprint
  {http://arxiv.org/abs/1810.02601} {arXiv:1810.02601 [physics.atom-ph]}
  \BibitemShut {NoStop}%
\bibitem [{\citenamefont {Flambaum}\ and\ \citenamefont
  {Kozlov}(2012)}]{FlambaumMolecule}%
  \BibitemOpen
  \bibfield  {author} {\bibinfo {author} {\bibfnamefont {V.~V.}\ \bibnamefont
  {Flambaum}}\ and\ \bibinfo {author} {\bibfnamefont {A.}~\bibnamefont
  {Kozlov}},\ }\href {\doibase 10.1103/PhysRevA.85.022505} {\bibfield
  {journal} {\bibinfo  {journal} {Phys. Rev. A}\ }\textbf {\bibinfo {volume}
  {85}},\ \bibinfo {pages} {022505} (\bibinfo {year} {2012})}\BibitemShut
  {NoStop}%
\bibitem [{\citenamefont {Brieger}(1984)}]{BRIEGER}%
  \BibitemOpen
  \bibfield  {author} {\bibinfo {author} {\bibfnamefont {M.}~\bibnamefont
  {Brieger}},\ }\href {\doibase https://doi.org/10.1016/0301-0104(84)85316-1}
  {\bibfield  {journal} {\bibinfo  {journal} {Chem. Phys.}\ }\textbf {\bibinfo
  {volume} {89}},\ \bibinfo {pages} {275 } (\bibinfo {year}
  {1984})}\BibitemShut {NoStop}%
\bibitem [{\citenamefont {Mizushima}\ \emph {et~al.}(1975)\citenamefont
  {Mizushima} \emph {et~al.}}]{mizushima1975theory}%
  \BibitemOpen
  \bibfield  {author} {\bibinfo {author} {\bibfnamefont {M.}~\bibnamefont
  {Mizushima}} \emph {et~al.},\ }\href@noop {} {\emph {\bibinfo {title} {Theory
  of rotating diatomic molecules}}}\ (\bibinfo  {publisher} {Wiley},\ \bibinfo
  {year} {1975})\BibitemShut {NoStop}%
\bibitem [{\citenamefont {Edmonds}(2016)}]{edmonds2016angular}%
  \BibitemOpen
  \bibfield  {author} {\bibinfo {author} {\bibfnamefont {A.~R.}\ \bibnamefont
  {Edmonds}},\ }\href@noop {} {\emph {\bibinfo {title} {Angular momentum in
  quantum mechanics}}}\ (\bibinfo  {publisher} {Princeton university press},\
  \bibinfo {year} {2016})\BibitemShut {NoStop}%
\bibitem [{\citenamefont {Grimaldi}\ \emph {et~al.}(1968)\citenamefont
  {Grimaldi}, \citenamefont {Lecourt},\ and\ \citenamefont
  {Moser}}]{DipoleLiHNaH}%
  \BibitemOpen
  \bibfield  {author} {\bibinfo {author} {\bibfnamefont {F.}~\bibnamefont
  {Grimaldi}}, \bibinfo {author} {\bibfnamefont {A.}~\bibnamefont {Lecourt}}, \
  and\ \bibinfo {author} {\bibfnamefont {C.}~\bibnamefont {Moser}},\ }\href
  {\doibase 10.1039/SF9680200059} {\bibfield  {journal} {\bibinfo  {journal}
  {Symp. Faraday Soc.}\ }\textbf {\bibinfo {volume} {2}},\ \bibinfo {pages}
  {59} (\bibinfo {year} {1968})}\BibitemShut {NoStop}%
\bibitem [{\citenamefont {Peterson}\ and\ \citenamefont
  {Woods}(1987)}]{DipoleBF}%
  \BibitemOpen
  \bibfield  {author} {\bibinfo {author} {\bibfnamefont {K.~A.}\ \bibnamefont
  {Peterson}}\ and\ \bibinfo {author} {\bibfnamefont {R.~C.}\ \bibnamefont
  {Woods}},\ }\href {\doibase 10.1063/1.452852} {\bibfield  {journal} {\bibinfo
   {journal} {J. Chem. Phys.}\ }\textbf {\bibinfo {volume} {87}},\ \bibinfo
  {pages} {4409} (\bibinfo {year} {1987})}\BibitemShut {NoStop}%
\bibitem [{\citenamefont {Childs}\ \emph {et~al.}(1986)\citenamefont {Childs},
  \citenamefont {Goodman},\ and\ \citenamefont {Goodman}}]{DipoleCaF}%
  \BibitemOpen
  \bibfield  {author} {\bibinfo {author} {\bibfnamefont {W.}~\bibnamefont
  {Childs}}, \bibinfo {author} {\bibfnamefont {G.}~\bibnamefont {Goodman}}, \
  and\ \bibinfo {author} {\bibfnamefont {L.}~\bibnamefont {Goodman}},\ }\href
  {\doibase https://doi.org/10.1016/0022-2852(86)90288-2} {\bibfield  {journal}
  {\bibinfo  {journal} {J. Mol. Spectrosc.}\ }\textbf {\bibinfo {volume}
  {115}},\ \bibinfo {pages} {215 } (\bibinfo {year} {1986})}\BibitemShut
  {NoStop}%
\bibitem [{NIS(2018)}]{NIST}%
  \BibitemOpen
  \href {https://webbook.nist.gov/chemistry/} {\enquote {\bibinfo {title}
  {{NIST} chemistry {WebBook}, {SRD} 69},}\ } (\bibinfo {year}
  {2018})\BibitemShut {NoStop}%
\bibitem [{\citenamefont {Akindinova}\ \emph {et~al.}(2010)\citenamefont
  {Akindinova}, \citenamefont {Chernov}, \citenamefont {Kretinin},\ and\
  \citenamefont {Zon}}]{Polarizability}%
  \BibitemOpen
  \bibfield  {author} {\bibinfo {author} {\bibfnamefont {E.~V.}\ \bibnamefont
  {Akindinova}}, \bibinfo {author} {\bibfnamefont {V.~E.}\ \bibnamefont
  {Chernov}}, \bibinfo {author} {\bibfnamefont {I.~Y.}\ \bibnamefont
  {Kretinin}}, \ and\ \bibinfo {author} {\bibfnamefont {B.~A.}\ \bibnamefont
  {Zon}},\ }\href {\doibase 10.1103/PhysRevA.81.042517} {\bibfield  {journal}
  {\bibinfo  {journal} {Phys. Rev. A}\ }\textbf {\bibinfo {volume} {81}},\
  \bibinfo {pages} {042517} (\bibinfo {year} {2010})}\BibitemShut {NoStop}%
\bibitem [{\citenamefont {Sutcliffe}(2007)}]{Sutcliffe}%
  \BibitemOpen
  \bibfield  {author} {\bibinfo {author} {\bibfnamefont {B.}~\bibnamefont
  {Sutcliffe}},\ }in\ \href {\doibase 10.1002/9780470141731.ch1} {\emph
  {\bibinfo {booktitle} {Advances in Chemical Physics}}}\ (\bibinfo
  {publisher} {Wiley-Blackwell},\ \bibinfo {year} {2007})\ pp.\ \bibinfo
  {pages} {1--121}\BibitemShut {NoStop}%
\bibitem [{\citenamefont {Baron}\ \emph {et~al.}(2014)\citenamefont {Baron}
  \emph {et~al.}}]{Baron:2013eja}%
  \BibitemOpen
  \bibfield  {author} {\bibinfo {author} {\bibfnamefont {J.}~\bibnamefont
  {Baron}} \emph {et~al.} (\bibinfo {collaboration} {ACME}),\ }\href {\doibase
  10.1126/science.1248213} {\bibfield  {journal} {\bibinfo  {journal}
  {Science}\ }\textbf {\bibinfo {volume} {343}},\ \bibinfo {pages} {269}
  (\bibinfo {year} {2014})},\ \Eprint {http://arxiv.org/abs/1310.7534}
  {arXiv:1310.7534 [physics.atom-ph]} \BibitemShut {NoStop}%
\bibitem [{\citenamefont {Eckel}\ \emph {et~al.}(2013)\citenamefont {Eckel},
  \citenamefont {Hamilton}, \citenamefont {Kirilov}, \citenamefont {Smith},\
  and\ \citenamefont {DeMille}}]{Eckel2013}%
  \BibitemOpen
  \bibfield  {author} {\bibinfo {author} {\bibfnamefont {S.}~\bibnamefont
  {Eckel}}, \bibinfo {author} {\bibfnamefont {P.}~\bibnamefont {Hamilton}},
  \bibinfo {author} {\bibfnamefont {E.}~\bibnamefont {Kirilov}}, \bibinfo
  {author} {\bibfnamefont {H.~W.}\ \bibnamefont {Smith}}, \ and\ \bibinfo
  {author} {\bibfnamefont {D.}~\bibnamefont {DeMille}},\ }\href {\doibase
  10.1103/PhysRevA.87.052130} {\bibfield  {journal} {\bibinfo  {journal} {Phys.
  Rev. A}\ }\textbf {\bibinfo {volume} {87}},\ \bibinfo {pages} {052130}
  (\bibinfo {year} {2013})}\BibitemShut {NoStop}%
\bibitem [{\citenamefont {Bickman}\ \emph {et~al.}(2009)\citenamefont
  {Bickman}, \citenamefont {Hamilton}, \citenamefont {Jiang},\ and\
  \citenamefont {DeMille}}]{Bickman2009}%
  \BibitemOpen
  \bibfield  {author} {\bibinfo {author} {\bibfnamefont {S.}~\bibnamefont
  {Bickman}}, \bibinfo {author} {\bibfnamefont {P.}~\bibnamefont {Hamilton}},
  \bibinfo {author} {\bibfnamefont {Y.}~\bibnamefont {Jiang}}, \ and\ \bibinfo
  {author} {\bibfnamefont {D.}~\bibnamefont {DeMille}},\ }\href {\doibase
  10.1103/PhysRevA.80.023418} {\bibfield  {journal} {\bibinfo  {journal} {Phys.
  Rev. A}\ }\textbf {\bibinfo {volume} {80}},\ \bibinfo {pages} {023418}
  (\bibinfo {year} {2009})}\BibitemShut {NoStop}%
\bibitem [{\citenamefont {Zaretskii}\ and\ \citenamefont
  {Lomonosov}(1979)}]{Zaretskii79}%
  \BibitemOpen
  \bibfield  {author} {\bibinfo {author} {\bibfnamefont {D.}~\bibnamefont
  {Zaretskii}}\ and\ \bibinfo {author} {\bibfnamefont {V.}~\bibnamefont
  {Lomonosov}},\ }\href@noop {} {\bibfield  {journal} {\bibinfo  {journal}
  {Pis’ma Zh. Eksp. Teor. Fiz.}\ }\textbf {\bibinfo {volume} {30}},\ \bibinfo
  {pages} {541} (\bibinfo {year} {1979})},\ \Eprint {http://arxiv.org/abs/[JETP
  Lett. 30, 508 (1979)]} {[JETP Lett. 30, 508 (1979)]} \BibitemShut {NoStop}%
\bibitem [{\citenamefont {Zaretskii}\ and\ \citenamefont
  {Lomonosov}(1981)}]{Zaretskii81}%
  \BibitemOpen
  \bibfield  {author} {\bibinfo {author} {\bibfnamefont {D.}~\bibnamefont
  {Zaretskii}}\ and\ \bibinfo {author} {\bibfnamefont {V.}~\bibnamefont
  {Lomonosov}},\ }\href@noop {} {\bibfield  {journal} {\bibinfo  {journal} {Zh.
  Eksp. Teor. Fiz.}\ }\textbf {\bibinfo {volume} {81}},\ \bibinfo {pages} {429}
  (\bibinfo {year} {1981})},\ \Eprint {http://arxiv.org/abs/[JETP 54, 229
  (1981)]} {[JETP 54, 229 (1981)]} \BibitemShut {NoStop}%
\bibitem [{\citenamefont {Dzyublik}(1992)}]{Dzyublik92}%
  \BibitemOpen
  \bibfield  {author} {\bibinfo {author} {\bibfnamefont {A.~Y.}\ \bibnamefont
  {Dzyublik}},\ }\href@noop {} {\bibfield  {journal} {\bibinfo  {journal} {Zh.
  Eksp. Teor. Fiz.}\ }\textbf {\bibinfo {volume} {102}},\ \bibinfo {pages}
  {120} (\bibinfo {year} {1992})},\ \Eprint {http://arxiv.org/abs/[JETP 75, 63
  (1992)]} {[JETP 75, 63 (1992)]} \BibitemShut {NoStop}%
\end{thebibliography}%

\end{document}